\documentclass{emulateapj}
\usepackage{apjfonts}
\usepackage{amsmath}
\slugcomment{accepted for publication in ApJ}
\shorttitle{GRB high-energy lags}
\shortauthors{Toma et al.}
\begin{document}
\title{
An up-scattered cocoon emission model of Gamma-Ray Burst high-energy lags
}
\author{Kenji Toma\altaffilmark{1,2}, 
Xue-Feng Wu\altaffilmark{1,2,3}, 
Peter M\'{e}sz\'{a}ros\altaffilmark{1,2,4}}

\altaffiltext{1}{Department of Astronomy and Astrophysics, Pennsylvania State 
University, 525 Davey Lab, University Park, PA 16802, USA}
\altaffiltext{2}{Center for Particle Astrophysics, Pennsylvania State University}
\altaffiltext{3}{Purple Mountain Observatory, Chinese Academy of Sciences, Nanjing 210008, China}
\altaffiltext{4}{Department of Physics, Pennsylvania State University}
\email{toma@astro.psu.edu}
\begin{abstract}
The {\it Fermi Gamma-ray Space Telescope} recently detected the most energetic 
gamma-ray burst so far, GRB 080916C, and reported its detailed temporal properties 
in an extremely
broad spectral range: (i) the time-resolved spectra are well described 
by broken power-law forms over the energy range of $10~{\rm keV}-10$~GeV, (ii) the
high-energy emission (at $\varepsilon > 100$~MeV) is delayed by $\approx 5$~s 
with respect to the $\varepsilon \lesssim 1$~MeV emission, and (iii) 
the emission onset times shift towards later times in higher energy bands.
We show that this behavior of the high-energy emission can be explained by a model
in which the prompt emission consists of two components: one is the emission component
peaking at $\varepsilon \sim 1$~MeV due to the synchrotron-self-Compton 
radiation of electrons accelerated in the internal shock of the jet and the other is 
the component peaking at $\varepsilon \sim 100$~MeV due to up-scattering of the 
photospheric X-ray emission of the expanding cocoon (i.e., the hot bubble produced by 
dissipation of the jet energy inside the progenitor star) off the same electrons 
in the jet.
Based on this model, we derive some constraints on the radius of the progenitor star
and the total energy and mass of the cocoon of this GRB, which may provide
information on the structure of the progenitor star and the physical conditions of 
the jet propagating in the star.
The up-scattered cocoon emission could be important for other Fermi GRBs as well.
We discuss some predictions of this model, including a prompt bright optical
emission and a soft X-ray excess.
\end{abstract}

\keywords{gamma rays: bursts --- gamma-rays: theory}

\section{Introduction}
\label{sec:intro}

Gamma-ray bursts (GRBs) were only sparsely observed in the $> 100$~MeV range, 
until the {\it Fermi} satellite was launched on June 11 2008 \citep{atwood09}.
Now {\it Fermi} provides extremely broad energy coverage, $8~{\rm keV}-300~{\rm GeV}$,
with high sensitivity for GRBs, and is accumulating a wealth of new data which
open a completely new window on the physics of GRBs.
The high-energy temporal and spectral data provided by {\it Fermi} can severely
constrain the physical parameters of the GRB emission region and the circumburst
environment, which will lead to a deeper understanding of the central engine and 
the GRB progenitors, and will also constrain models of high-energy cosmic ray 
acceleration \citep[for recent reviews,][]{falcone08,zhang07,meszaros06}.

The GRB detected by {\it Fermi} on September 16 2008 (GRB 080916C) shows
several important new properties \citep{abdo09}:
\begin{itemize}
\item[(i)]
The time-resolved spectra (with resolution $\sim 5-50$~s) are well fitted by an 
empirical broken-power-law function \citep[the so-called Band function,][]{band93} 
from 8~keV up to a photon with energy $\approx 13.2$~GeV.
\item[(ii)]
The $\varepsilon > 100$~MeV emission is not detected together with the first 
$\varepsilon \lesssim 1$~MeV pulse and the onset of the $\varepsilon>100$~MeV 
emission coincides with the rise of the second pulse ($\approx 5$~s after the trigger).
\item[(iii)]
Most of the emission in the second pulse shifts towards later times as higher 
energies are considered.
\item[(iv)]
The $\varepsilon>100$~MeV emission lasts at least 1400~s, while photons with 
$\varepsilon<100$~MeV are not detected past 200~s.
\item[(v)]
The redshift $z \simeq 4.35$ \citep{greiner09} and the fluence $\approx 2.4 \times
10^{-4}~{\rm erg}~{\rm cm}^{-2}$ in the 10~keV - 10~GeV range
mean that this is the largest reported isotropic
$\gamma$-ray energy release, $E_{\gamma,{\rm iso}} \simeq 8.8\times 10^{54}$~erg.
\end{itemize}
Some other bursts detected by the LAT detector of {\it Fermi} also display
high-energy lags, similar to the properties (ii) and/or (iii) \citep{abdo09}, 
and then they should be very important to understand the prompt emission mechanism
of GRBs.
We will call the $\varepsilon \lesssim 1$~MeV emission and 
the $\varepsilon>100$~MeV emission "MeV emission" and "high-energy emission", 
respectively.

A simple physical picture for the property (i) is that the prompt emission consists of 
a single emission component, such as synchrotron radiation of electrons accelerated
in internal shocks of a relativistic jet.
In this picture, the peak of the MeV pulse could be attributed to the cessation of
the emission production (i.e., the shock crossing of the shell) and the 
subsequent emission could come from the high latitude regions of the shell
\citep[e.g.,][]{dermer04,zhang06}.
Thus the observed high-energy lag for the second pulse (property (iii)) requires
that the electron energy spectrum should be harder systematically in the higher
latitude region.
This would imply that the particle acceleration process should definitely depend on 
the global parameters of the jet, e.g., the angle-dependent relative Lorentz factor 
of the colliding shells, but such a theory has not been formulated yet.
The property (ii) could be just due to the fact that the two pulses originate in
two internal shocks with different physical conditions for which the electron
energy spectrum of the second internal shock is harder than that of the first one
\citep{abdo09}.

Another picture is that the prompt emission consists of the MeV component and 
a delayed high-energy component.
The latter component could be produced by hadronic effects (i.e., photo-pion
process and proton synchrotron emission) \cite[e.g.,][]{dermer02,asano08}, 
because the acceleration of protons or ions to high energies may be delayed 
behind the electron acceleration.
However, it is not clear that hadronic effects can reproduce the smooth Band
spectrum \citep[see also][]{wang09}.

\begin{figure}
\plotone{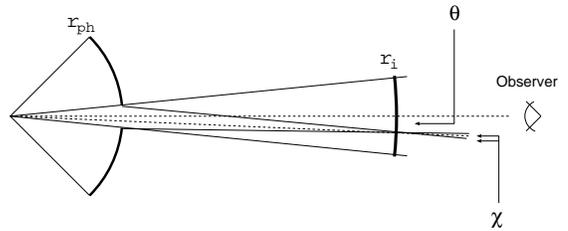}
\caption{
Geometry of our model.
The jet with opening angle $\theta_j \simeq 0.01$ and the cocoon with 
opening angle $\theta_c \simeq 0.8$ are ejected from the collapsing star.
The cocoon becomes optically thin at 
$r=r_{\rm ph} \simeq 2.1 \times 
10^{14}~{\rm cm}~E_{c,52}^{1/2}~(\Gamma_c/50)^{-1/2}$ 
some time ($\lesssim 5$~s) after the burst trigger (see text for detail).
The second pulse of the prompt emission is produced by the accelerated electrons
in the internal shock of the jet at $r=r_i \simeq 2.2 \times 10^{16}~{\rm cm}~
\Gamma_{j,3}^2~(\Delta t_i/2~{\rm s})$, and the cocoon X-ray photons
are up-scattered by the same electrons at $r=r_i$ into the GeV range.
The jet is assumed to be observed from the direction of the jet axis,
and $\theta$ describes the emitting point of the high-latitude emission.
The angle $\chi$ describes the difference of the directions of the
expansion and the cocoon photon beams, which can be neglected for our
adopted parameters.
}
\label{fig:geometry}
\end{figure}

In this paper, we discuss a different two-component emission picture in which
the delayed high-energy component is produced by leptonic process 
(i.e., electron inverse Compton scattering)
\footnote{
Other ideas for the delayed high-energy emission of GRB 080916C
in the leptonic model are proposed in \citet{li09}, \citet{zou09}, and \citet{fan09}
\citep[see also][]{aoi09}.
}.
We focus on the effect that the ambient radiation up-scattered by the accelerated
electrons in the jet can have a later peak than that of the synchrotron and 
synchrotron-self-Compton (SSC) emission of the same electrons
(corresponding to the property (iii))
\citep{wang06,fan08,beloborodov05}.
Provided that the seed photons for the Compton scattering come from the region
behind the electron acceleration region of the jet (see Figure~\ref{fig:geometry}), 
the up-scattered high-energy photon field is highly anisotropic in the comoving
frame of the jet, i.e., the emissivity is much larger for the head-on collisions of 
the electrons and the seed photons. 
As a result, a stronger emission is observed from the higher latitude regions,
and thus its flux peak lags behind the synchrotron and SSC emission.

Here we propose that the seed photons may be provided by the photospheric 
emission of an expanding cocoon.
GRB 080916C is a long GRB, and it may originate from the collapse of a massive 
star.
The relativistic jet produced in the central region penetrates the star and deposits
most of its energy output into a thermal bubble, or cocoon,
until it breaks out of the star \citep[e.g.,][]{meszaros01,wzhang03,mizuta06}.
The cocoon can store an energy comparable or larger than the energy of
the prompt emission of the jet, and thus it may make an observable signature 
outside the star \citep{ramirez02,peer06}.
The cocoon escaping from the star will emit soft X-rays, and these can be 
up-scattered by the accelerated electrons in the jet into the high-energy range.
The optical thinning of the expanding cocoon may be delayed behind the prompt
emission of the jet, so that the onset of the high-energy emission is delayed
behind the MeV emission (corresponding to the property (ii)).
Thus this model has the potential for explaining the two delay timescales;
the delayed onset of the high-energy photons (property (ii))
is due to the delayed emission of the cocoon,
while the high-energy lag within the second pulse (property (iii)) is due to
the anisotropic inverse Compton scattering.
We also show that the combination of the time-averaged spectra of the SSC and
the up-scattered cocoon (UC) emission is roughly consistent with the observed
smooth power-law spectrum (property (i)) (see Figure~\ref{fig:spectrum}).

As we will explain below, the UC emission is short-lived and may not
account for the whole high-energy emission, which lasts much longer than
the MeV emission (property (iv)).
It is natural that the high-energy emission in later times is related to
the afterglow, i.e., produced by the external shock which propagates in the 
circumburst medium. This possibility is studied by \citet{kumar09b}.
They claim that even the onset phase of the high-energy emission is produced
by the external shock. However, the rise of the flux of GRB 080916C in the LAT
energy range ($\sim t^6$) is too steep for the external shock to reproduce it.
Thus at least the first part of the delayed high-energy component 
of this burst should be related to the prompt emission.

Although some {\it Fermi}-LAT GRBs show  high-energy lags, the 
observational analysis of the delayed high-energy emission is fairly
complete only for GRB 080916C at the moment, and its data is quite extensive.
Thus we focus on this burst to examine whether the UC emission is viable 
for its properties in the high energy range in this paper.
However, our modeling is generic, and we will show that the UC emission
could be important for some other {\it Fermi}-LAT GRBs (including short GRBs).

This paper is organized as follows.
In \S~\ref{sec:cocoon_dynamics}, we derive the delay time of the onset of 
the cocoon photospheric emission behind the MeV pulse of the jet,
and calculate the flux of the cocoon emission. 
In \S~\ref{sec:emission_formalism}, we constrain the physical conditions of the 
emitting region of the jet by assuming that the MeV emission is due to the 
1st-order SSC radiation, and show that the UC flux of the jet and its lag timescale 
are compatible with the observed GeV emission.
In \S~\ref{sec:fitting}, we calculate the spectrum and lightcurve in a simple
numerical manner and find appropriate values of the physical parameters of
the cocoon and jet for reproducing the observed spectrum at 
$\varepsilon \gtrsim 1$~MeV and lightcurve of GRB 080916C as the combination of 
the 1st-order SSC and UC emission.
We discuss the emission at $\varepsilon < 1$~MeV
and the thermal emission of the jet in \S~\ref{sec:discussion}.
The UC emission for other long and short GRBs are discussed in \S~\ref{sec:other}.
We summarize our model and its implications in \S~\ref{sec:summary}.
We use ${\rm keV}~{\rm cm}^{-2}~{\rm s}^{-1}~{\rm keV}^{-1}$ and
${\rm keV}~{\rm cm}^{-2}~{\rm s}^{-1}$ for the units of the flux density
and flux, respectively, for comparison with Fermi papers, e.g., \citet{abdo09}.
The conversion factors to usual units are useful: 
$1~{\rm keV}~{\rm cm}^{-2}~{\rm s}^{-1}~{\rm keV}^{-1} = 
6.63 \times 10^{-27}~{\rm erg}~{\rm cm}^{-2}~{\rm s}^{-1}~{\rm Hz}^{-1}$ 
and $1~{\rm keV}~{\rm cm}^{-2}~{\rm s}^{-1} = 
1.60 \times 10^{-9}~{\rm erg}~{\rm cm}^{-2}~{\rm s}^{-1}$.

\section{Cocoon dynamics}
\label{sec:cocoon_dynamics}

The dynamics of the cocoon after escaping from the star is studied in 
\citet{ramirez02} and \citet{peer06}.
Based on these studies, we derive the delay time of the cocoon emission behind
the prompt emission of the jet and the energy flux of the cocoon emission.

\subsection{Time delay of the cocoon emission onset}

The cocoon dynamics is dominated by its radiation energy while it is inside the star.
We take the total energy and the total mass of the cocoon and the stellar radius
as parameters, $E_c$, $M_c$, and $r_*$, respectively.
After the jet breaks out of the star, the cocoon expands with the sound speed,
$c_s = c/\sqrt{3}$, where $c$ is the speed of light.
The cocoon expands by its own thermal pressure in the comoving frame 
as expected in the standard fireball model \citep{meszaros93,piran93}.
Its expansion speed in the comoving frame suddenly becomes close to $c$, and then
the opening angle of the cocoon measured in the central engine frame is given by
$\theta_c \simeq \Gamma_{c,{\rm ex}}^{-1}$, where $\Gamma_{c,{\rm ex}}$ is the 
Lorentz factor corresponding to $c_s$.
Thus we obtain $\theta_c \simeq \sqrt{2/3} \simeq 0.8$.
\citet{morsony07} have performed detailed numerical simulations of the 
cocoon expansion outside the star, which show that the opening angle of the 
expanding cocoon is $\sim 40^{\circ}$. This is consistent with the above
analytical estimate.

The cocoon material accelerates as $\Gamma \propto r$ and 
reaches the terminal Lorentz factor $\Gamma_c = E_c/M_c c^2$ at
\begin{equation}
r_s = r_* \Gamma_c = 5.0\times10^{12}~{\rm cm}~r_{*,11} 
\left(\frac{\Gamma_c}{50}\right).
\label{eq:saturation_radius}
\end{equation}
Hereafter we adopt the notation $Q = 10^x Q_x$ in cgs units.
For $r>r_s$, the cocoon material is dominated by the kinetic energy.
The radiation stored in the cocoon is released when the opacity for the electron
scattering becomes less than unity.
The photosphere radius is given by
\begin{equation}
r_{\rm ph} \simeq 
\left[\frac{E_c \sigma_T}{2\pi (1-\cos\theta_c) \Gamma_c m_p c^2}\right]^{1/2}
\simeq 2.1 \times 10^{14}~{\rm cm}~E_{c,52}^{1/2} 
\left(\frac{\Gamma_c}{50}\right)^{-1/2},
\label{eq:photospheric_radius}
\end{equation}
where $\sigma_T$ is Thomson cross section.

The cocoon may become optically thin later than the onset of the first MeV pulse
(i.e., the burst trigger time).
Let $t=0$ be the photon arrival time at the earth emitted at the stellar surface
at the jet breakout.
The first MeV pulse produced within the jet at $r=r_i$ is observed at
\begin{equation}
t \simeq \Delta t_i \equiv \frac{r_i}{2 c \Gamma_j^2} (1+z),
\end{equation}
where $\Gamma_j$ is the bulk Lorentz factor of the jet and $z$ is the source 
redshift.
This timescale is comparable to the angular spreading timescale of the pulse,
and we can take $\Delta t_i \simeq 2$~s for GRB 080916C.
The second MeV pulse is observed $\simeq 5$~s after the burst trigger, i.e.,
at $t \simeq \Delta t_i + 5 \simeq 7$~s.
On the other hand, the cocoon photospheric emission is observed at
\begin{equation}
t \simeq \Delta t_c \equiv \frac{r_{\rm ph}}{2c \Gamma_c^2} (1+z)
\simeq 7.5~{\rm s}~E_{c,52}^{1/2} \left(\frac{\Gamma_c}{50}\right)^{-5/2},
\label{eq:delta_t_c}
\end{equation}
where we have used $z \simeq 4.35$.
The cocoon photospheric emission may be observed from $t=\Delta t_c$ to 
$t = \Delta t_c + \Delta t_d$, where $\Delta t_d \sim \Delta t_c$ is the time during
which the cocoon will be adiabatically cooled.
If internal dissipation occurs in the jet at $r = r_i > r_{\rm ph}$ making the 
second MeV pulse within the duration of the cocoon emission, 
the cocoon photons may be up-scattered to higher energies by the energetic 
electrons within the dissipation region of the jet, 
which may be observed along with the second MeV pulse.
Therefore we require a condition $\Delta t_i < \Delta t_c \lesssim \Delta t_i + 5$.
This condition puts a constraint to the physical parameters of the cocoon,
\begin{equation}
0.3 < E_{c,52}^{1/2} \left(\frac{\Gamma_c}{50}\right)^{-5/2} \lesssim 0.9
\label{eq:co_constraint1}
\end{equation}
We adopt the parameters 
$E_c \simeq 10^{52}$~erg and $\Gamma_c \simeq 50$
for the purposes of calculating the flux of the cocoon emission.

\subsection{Flux of the cocoon emission}

The kinetic energy of the cocoon can be dissipated into radiation energy, 
e.g., via internal shocks or magnetic reconnection during the expansion.
Since the cocoon material may initially have some inhomogeneities caused by the 
interaction with the jet and the stellar envelope, on scales $\sim \alpha r_*$,
where $\alpha \leq 1$ is a numerical factor, internal shocks may occur at
\citep{rees94,meszaros00}
\begin{equation}
r_d \simeq 2 \alpha r_* \Gamma_c^2 \simeq 5.0 \times 10^{13}~{\rm cm}~
r_{*,11} \left(\frac{\Gamma_c}{50}\right)^2 \left(\frac{\alpha}{0.1}\right).
\end{equation}
We focus on the case of $r_s < r_d < r_{\rm ph}$, in which
the optical depth for the electron scattering in the 
dissipation region is large and the photons produced by the dissipation
undergo multiple Compton scattering before escape.
As a result, the emerging spectrum is expected to be quasi-thermal \citep{peer06}.
The condition $r_s < r_d < r_{\rm ph}$ reduces, 
by equations (\ref{eq:saturation_radius}), (\ref{eq:photospheric_radius}), and
(\ref{eq:co_constraint1})
into
\begin{equation}
0.01~\left(\frac{\Gamma_c}{50}\right)^{-1} < \alpha < 0.4~r_{*,11}^{-1}.
\label{eq:co_constraint2}
\end{equation}

The comoving temperature of the cocoon when its opening angle becomes $\theta_c$ 
is approximately given by
\footnote{
The initial temperature of the cocoon may be estimated more accurately.
The temperature of the cocoon just before the jet break-out is 
$T'_i = [E_c/(V'_i a)]^{1/4}$, where the volume of the cocoon can be
approximated as $V'_i \approx \pi \theta_{ci}^2 r_*^3/3$ ($\theta_{ci}$
is the opening angle of the cocoon just before the jet break-out.)
The cocoon escapes from the star and its comoving volume is
$V'_{\rm init} \approx 2\pi(1-\cos\theta_c) r_*^3/3$ when the opening angle
becomes $\theta_c$. (The relativistic contraction effect is negligible.)
At that time the temperature is calculated as 
$T'_{\rm ex} \simeq T'_i (V'_{\rm init}/V'_i)^{-1/3} \simeq 0.93~T'_{\rm init} \theta_{ci,-1}^{1/6}.$
To simplify the equations in the following, we neglect the weak dependence on 
$\theta_{ci}$ and adopt $T'_{\rm init}$.}
\begin{equation} 
T'_{\rm init} \simeq \left[\frac{E_c}{2\pi (1-\cos\theta_c) r_*^3 a}\right]^{1/4}
\simeq 1.6 \times 10^8~{\rm K}~E_{c,52}^{1/4} r_{*,11}^{-3/4},
\end{equation}
where $a$ is the Stefan constant.
The comoving temperature evolves adiabatically as $T' \propto r^{-1}$ 
for $r_* < r < r_s$ (i.e., in the radiation-dominated phase) and 
$T' \propto r^{-2/3}$ at $r > r_s$ (i.e., in the matter-dominated phase)
\citep{meszaros93}, and thus the temperature at $r=r_d$ is 
\begin{equation}
T'_{ad} = T'_{\rm init} \left(\frac{r_s}{r_*}\right)^{-1}
\left(\frac{r_d}{r_s}\right)^{-2/3}.
\end{equation}
If the fraction $\epsilon_d$ of the kinetic energy is converted into thermal
energy at $r=r_d$, the corresponding temperature is 
\begin{equation}
T'_d = \left[\frac{E_c \epsilon_d / \Gamma_c}
{2\pi (1-\cos\theta_c) r_d^2 r_* \Gamma_c a}\right]^{1/4}
= T'_{ad} \left(\frac{r_d}{r_s}\right)^{1/6} \epsilon_d^{1/4}.
\end{equation}
If the dissipation is not too strong, i.e., $\epsilon_d \lesssim (r_d/r_s)^{-2/3}$,
the temperature at $r = r_d$ is given by $T'_{ad}$ \citep[cf.][]{rees05}.
After the dissipation the temperature evolves as $T' \propto r^{-2/3}$ since
the cocoon shell is still dominated by the kinetic energy and the shell
spreading effect is negligible up to $r \sim 2r_* \Gamma_c^2 > r_{\rm ph}$.
The comoving temperature at the photosphere radius is then estimated by
\begin{equation}
T'_{\rm ph} = T'_{ad} \left(\frac{r_{\rm ph}}{r_d}\right)^{-2/3}
\simeq 2.7 \times 10^5~{\rm K}~E_{c,52}^{-1/12} r_{*,11}^{-1/12}.
\end{equation}

The photospheric emission of the cocoon is expected to have a quasi-thermal spectrum
\begin{equation}
F^{\rm co}_{\varepsilon} = F^{\rm co}_{\varepsilon_{\rm ph}} \times \left\{
\begin{array}{lc}
\left(\frac{\varepsilon}{\varepsilon^{\rm co}_{\rm ph}}\right)^2  
& {\rm for}~\varepsilon < \varepsilon^{\rm co}_{\rm ph}, \\
\left(\frac{\varepsilon}{\varepsilon^{\rm co}_{\rm ph}}\right)^{\beta} 
& {\rm for}~ \varepsilon^{\rm co}_{\rm ph} < \varepsilon < 
\varepsilon^{\rm co}_{\rm cut},
\end{array}
\right.
\label{eq:cocoon_spectrum}
\end{equation}
where $\varepsilon_{\rm ph}^{\rm co}$ and $F_{\varepsilon_{\rm ph}}^{\rm co}$ are 
given by
\begin{equation}
\varepsilon_{\rm ph}^{\rm co} \simeq 2.82~ kT'_{\rm ph} \frac{2\Gamma_c}{1+z}
\simeq 1.2~{\rm keV}~E_{c,52}^{-1/12} r_{*,11}^{-1/12} 
\left(\frac{\Gamma_c}{50}\right)
\label{eq:cocoon_nu}
\end{equation}
\begin{eqnarray}
F_{\varepsilon_{\rm ph}}^{\rm co} & \simeq & \frac{(1+z)^3}{d_L^2} 
\frac{2\pi (\nu_{\rm ph}^{\rm co})^2}{c^2} kT'_{\rm ph} 
\Gamma_c \left(\frac{r_{\rm ph}}{\Gamma_c}\right)^2 \nonumber \\
& \simeq & 31~{\rm keV}~{\rm cm}^{-2}~{\rm s}^{-1}~{\rm keV}^{-1}~
E_{c,52}^{3/4} r_{*,11}^{-1/4},
\label{eq:cocoon_fnu}
\end{eqnarray}
where $\nu_{\rm ph}^{\rm co} = \varepsilon_{\rm ph}^{\rm co}/h$ ($h$ is the Planck
constant), and we have taken the luminosity distance of GRB 080916C as 
$d_L \simeq 1.2 \times 10^{29}$~cm.
These are derived by assuming that the cocoon is viewed on-axis, 
and this assumption is validated in \S~\ref{sec:jet_overall}.
\citet{peer06} carried out detailed numerical calculations of the radiative processes
in the cocoon and show that the cocoon photospheric emission has the above
spectral form, with $\beta \sim -1$ regardless of the dissipation process if
$r_d < r_{\rm ph}$.
The cutoff energy $\varepsilon_{\rm cut}^{\rm co}$ is given by 
$\sim 30 \times \varepsilon_{\rm ph}^{\rm co}$.

The observation of GRB 080916C shows that there is no excess from the Band spectrum
at the X-ray band, $\gtrsim 10$~keV, and we obtain a rough upper limit of the
cocoon X-ray emission 
$\varepsilon_{\rm ph}^{\rm co} F_{\varepsilon_{\rm ph}}^{\rm co} \lesssim
40~{\rm keV}~{\rm cm}^{-2}~{\rm s}^{-1}$.
This limit leads to another constraint on the cocoon parameters,
\begin{equation}
r_{*,11} \gtrsim 0.8~E_{c,52}^2 \left(\frac{\Gamma_c}{50}\right)^3.
\label{eq:co_constraint3}
\end{equation}

\section{Up-scattered cocoon (UC), synchrotron, and SSC emission}
\label{sec:emission_formalism}

We describe here the calculations of the fluxes of the UC, synchrotron,
and SSC emission of the jet for the second pulse of GRB 080916C 
in the internal shock model.
The {\it Fermi} observation allows us to constrain some global parameters of 
the jet (\S~\ref{sec:jet_overall}), and we show that the 1st-order SSC radiation of 
electrons accelerated in an internal shock can explain the peak energy
and the peak flux of the MeV emission (\S~\ref{sec:is_model}).
Then we show that the same electrons can produce the UC emission which lags
behind the SSC emission and has the peak energy and the peak flux compatible with
the observation in the GeV range (\S~\ref{sec:aic}).

\subsection{Overall properties of the jet}
\label{sec:jet_overall}

We can constrain, from the {\it Fermi} observation, the global physical parameters of the 
jet: the bulk Lorentz factor $\Gamma_j$, the emission radius of the second pulse $r_i$, 
and the opening angle $\theta_j$.
First of all, from the absence of a $\gamma\gamma$ absorption cutoff,
we obtain a lower limit on the bulk Lorentz factor of the jet, 
$\Gamma_j \gtrsim 870$ \citep{abdo09}.
Since the angular spreading timescale of the pulse is $\Delta t_i \simeq 2$~s
for the second pulse, similar to the first pulse, the emission radius is
estimated by
\begin{equation}
r_i \simeq 2c \Gamma_j^2 \frac{\Delta t_i}{1+z} 
\simeq 2.2 \times 10^{16}~{\rm cm}~\Gamma_{j,3}^2 
\left(\frac{\Delta t}{2~{\rm s}}\right).
\end{equation}

Since GRB 080916C is so bright, it is probable that the jet is viewed on-axis,
and we adopt this assumption as a simplification (see Figure~\ref{fig:geometry}).
In this case, the cocoon is viewed off-axis, since the jet cone will not be
filled with the cocoon material.
The cocoon emission is thus less beamed, but this off-axis effect is not
significant because the opening angle of the jet can be estimated to be small.
The isotropic $\gamma$-ray energy of this burst is $8.8 \times 10^{54}$~erg.
To obtain a realistic value of the collimation-corrected $\gamma$-ray energy,
$\lesssim 10^{51}$~erg, the jet opening angle is constrained by 
$\theta_j \lesssim 0.015$.
We adopt $\theta_j \simeq 0.01$, and having adopted a nominal value of 
$\Gamma_c \simeq 50$ in accord with the observed time delay of the high-energy emission 
(equation \ref{eq:co_constraint1}), we obtain
\begin{equation}
\Gamma_c \theta_j \simeq 0.5 < 1.
\end{equation}
Thus the off-axis dimming and softening effects are not significant for
the cocoon emission.

\subsection{Internal shock model for the MeV emission}
\label{sec:is_model}

We assume that the jet is dominated by the kinetic energy of protons (instead of 
magnetic energy) and we estimate the physical parameters 
of the jet dissipation region for the second MeV pulse.
In the scenario where the dissipation is due to internal shocks \citep{rees94}
\citep[see also][]{panaitescu00,guetta03,gupta07,kumar08,bosnjak09},
the collisionless shock waves can amplify the magnetic field and 
accelerate electrons to a power-law energy distribution, which then 
produce synchrotron radiation and SSC radiation.
We find that the 1st-order SSC radiation can account for the observed
MeV emission.
At $r=r_i$, the comoving number density of the jet is estimated by
\begin{eqnarray}
n' &=& \frac{L \Delta t_i/(1+z)}{4\pi r_i^3 m_p c^2}
= \frac{L(1+z)^2}{32\pi m_p c^5 \Gamma_j^6 \Delta t_i^2}  \nonumber \\
&\simeq& 1.7 \times 10^7~{\rm cm}^{-3}~L_{55} \Gamma_{j,3}^{-6}
\left(\frac{\Delta t_i}{2~{\rm s}}\right)^{-2},
\end{eqnarray}
where $L$ is the isotropic-equivalent luminosity of the jet, and
we have used the fact that the dynamical timescale of the emitting shell
measured in the central engine frame 
is comparable to the angular spreading timescale given by $\Delta t_i/(1+z)$.
The optical depth of the electron scattering is calculated as
\begin{equation}
\tau = \sigma_T n' \frac{r_i}{\Gamma_j} \simeq 2.6 \times 10^{-4}~
L_{55} \Gamma_{j,3}^{-5} \left(\frac{\Delta t_i}{2~{\rm s}}\right)^{-1}.
\end{equation}
This indicates that the dissipation region is optically thin, which
allows the non-thermal synchrotron and SSC emission to be observed.

The internal energy density produced by the internal shock is given by
$u' = n' \theta_p m_p c^2$, where $\theta_p$ is a mean random Lorentz factor of the
shocked protons and it is of order unity.
Assuming that a fraction $\epsilon_B$ of the internal energy of the protons 
is carried into the magnetic field, the field strength is estimated by
\begin{equation}
B' = (8\pi \epsilon_B u')^{1/2} \simeq 2.6~{\rm G}~L_{55}^{1/2}
\Gamma_{j,3}^{-3} \left(\frac{\Delta t_i}{2~{\rm s}}\right)^{-1} \theta_p^{1/2}
\left(\frac{\epsilon_B}{10^{-5}}\right)^{1/2}.
\end{equation}
Assuming that a fraction $\epsilon_e$ of the internal energy of the protons is
given to the electrons, the minimum Lorentz factor of the electrons is
given by
\begin{equation}
\gamma_m = \frac{p-2}{p-1} \frac{m_p}{m_e} \epsilon_e \theta_p
\simeq 400~\theta_p \left(\frac{\epsilon_e}{0.4}\right),
\end{equation}
where $p$ is the index of the electron energy distribution and we have assumed 
$p \simeq 3.2$ (see \S~\ref{sec:fitting} for this number).
We find that the electrons are radiatively cooled significantly within
the dynamical time for the parameters for fitting the observational data
(see \S~\ref{sec:fitting}), 
and the electron energy spectrum averaged over the dynamical time is expected to be
\begin{equation}
N'(\gamma) \propto \left\{
\begin{array}{lc}
\gamma^{-2} & {\rm for}~\gamma_c < \gamma < \gamma_m, \\
\gamma^{-p-1} & {\rm for}~\gamma_m < \gamma,
\end{array}
\right.
\label{eq:electron_spectrum}
\end{equation}
where $\gamma_c$ is the cooling Lorentz factor of the electrons, estimated by
\begin{equation}
\gamma_c = \frac{6\pi m_e c}{\sigma_T {B'}^2 (1+x_1+x_1 x_{\rm uc}) 
r_i/(c\Gamma_j)}.
\label{eq:gamma_c1}
\end{equation}
Here $x_1$ and $x_{\rm uc}$ are the luminosity
ratios of the 1st-order SSC radiation to the synchrotron radiation and the
UC radiation to the 1st-order SSC radiation, respectively.
The 2nd-order SSC radiation luminosity is negligible compared to the 1st-order
SSC radiation because of the Klein-Nishina suppression.
Since $x_1$ and $x_{\rm uc}$ are complicated functions of $\gamma_c$, one cannot
obtain a simple analytical formula for estimating $\gamma_c$, but we can estimate
$\gamma_c$ for the purpose of fitting the data of GRB 080916C in this model.
We find $\gamma_c \sim \gamma_m$ for fitting the observational data 
(see \S~\ref{sec:fitting}), and thus the parameter $x_1$ is written as
\begin{equation}
x_1 = \frac{4}{3} \tau \gamma_c^2 \frac{p}{p-2}.
\label{eq:gamma_c2}
\end{equation}
We find that $x_{\rm uc} \sim 0.5$ in order for the UC emission flux to be 
compatible with the observed flux in the GeV range, and thus
equation (\ref{eq:gamma_c1}) and (\ref{eq:gamma_c2}) lead to a formula for 
estimating the cooling Lorentz factor,
\begin{equation}
\gamma_c \simeq 360~
\left(\frac{\epsilon_B}{10^{-5}}\right)^{-1/3} 
\left(\frac{\tau}{4\times10^{-4}}\right)^{-2/3}.
\end{equation}

The synchrotron characteristic energy and the synchrotron peak flux 
(at the synchrotron energy corresponding to $\gamma_c$) are estimated by
\begin{eqnarray}
\varepsilon_m &\simeq& \frac{3h e B'}{4\pi m_e c}\gamma_m^2 \frac{2\Gamma_j}{1+z} 
\nonumber \\
&\simeq& 2.7~{\rm eV}~L_{55}^{1/2} \Gamma_{j,3}^{-2} 
\left(\frac{\Delta t_i}{2~{\rm s}}\right)^{-1} \theta_p^{5/2} 
\left(\frac{\epsilon_B}{10^{-5}}\right)^{1/2} \left(\frac{\epsilon_e}{0.4}\right)^2,
\label{eq:syn_num}
\end{eqnarray}
\begin{eqnarray}
F_{\varepsilon_c} &\simeq& \frac{\sqrt{3} e^3 B' N}{m_e c^2} 
\frac{2\Gamma_j (1+z)}{4\pi d_L^2} \nonumber \\
&\simeq& 1.3 \times 10^4~{\rm keV}~{\rm cm}^{-2}~{\rm s}^{-1}~{\rm keV}^{-1}~
L_{55}^{3/2} \Gamma_{j,3}^{-3} \theta_p^{1/2} 
\left(\frac{\epsilon_B}{10^{-5}}\right)^{1/2},
\label{eq:syn_fnuc}
\end{eqnarray}
where $N = [L \Delta t_i/(1+z)]/(\Gamma_j m_p c^2)$.
The 1st-order SSC characteristic energy and the SSC peak flux are approximately
\begin{eqnarray}
\varepsilon_m^{\rm SC} &\simeq& 4\gamma_m^2 \varepsilon_m \nonumber \\ &\simeq&
1.7~{\rm MeV}~L_{55}^{1/2} \Gamma_{j,3}^{-2} 
\left(\frac{\Delta t_i}{2~{\rm s}}\right)^{-1} \theta_p^{9/2} 
\left(\frac{\epsilon_B}{10^{-5}}\right)^{1/2} \left(\frac{\epsilon_e}{0.4}\right)^4,
\label{eq:SSC_num}
\end{eqnarray}
\begin{eqnarray}
F_{\varepsilon_c}^{\rm SC} &\simeq& \tau F_{\varepsilon_c} \nonumber \\
&\simeq& 3.4~{\rm keV}~{\rm cm}^{-2}~{\rm s}^{-1}~{\rm keV}^{-1}~L_{55}^{5/2}
\Gamma_{j,3}^{-8} \left(\frac{\Delta t_i}{2~{\rm s}}\right)^{-1} \theta_p^{1/2}
\left(\frac{\epsilon_B}{10^{-5}}\right)^{1/2}.
\label{eq:SSC_fnuc}
\end{eqnarray}

The synchrotron self-absorption optical depth at $\varepsilon < \varepsilon_c 
\equiv \varepsilon_m (\gamma_c/\gamma_m)^2$ can be estimated by 
\citep{matsumiya03,toma08}
\begin{equation}
\tau_a \simeq \frac{2\pi^2 (p+3)p}{9 \Gamma_E (\frac{1}{3}) (p+\frac{5}{3})}
\frac{e}{\sigma_T} \tau {B'}^{-1} \gamma_c^{-5} 
\left(\frac{\varepsilon}{\varepsilon_c}\right)^{-5/3},
\end{equation}
where $\Gamma_E(x)$ is the Euler's Gamma function and we have assumed that 
$\gamma_c \sim \gamma_m$.
Then the ratio of the frequency at which $\tau_a = 1$ to $\varepsilon_c$ is given by
\begin{equation}
\frac{\varepsilon_a}{\varepsilon_c} \simeq 0.20~L_{55}^{3/10} \Gamma_{j,3}^{-6/5} 
\theta_p^{-3/10} \left(\frac{\epsilon_B}{10^{-5}}\right)^{-3/10} 
\left(\frac{\gamma_c}{400}\right)^{-3}.
\label{eq:self_absorption}
\end{equation}

The spectrum of the 1st-order SSC emission is described by
\begin{equation}
F_{\varepsilon} \simeq F_{\varepsilon_c}^{\rm SC} \times \left\{
\begin{array}{lc}
\left(\frac{\varepsilon_a^{\rm SC}}{\varepsilon_c^{\rm SC}}\right)^{1/3} 
\left(\frac{\varepsilon}{\varepsilon_a^{\rm SC}}\right)^{1}, 
& {\rm for}~\varepsilon<\varepsilon_a^{\rm SC}, \\
\left(\frac{\varepsilon}{\varepsilon_c^{\rm SC}}\right)^{1/3},
& {\rm for}~\varepsilon_a^{\rm SC} < \varepsilon < \varepsilon_c^{\rm SC}, \\
\left(\frac{\varepsilon}{\varepsilon_c^{\rm SC}}\right)^{-1/2},
& {\rm for}~\varepsilon_c^{\rm SC}<\varepsilon<\varepsilon_m^{\rm SC}, \\
\left(\frac{\varepsilon_m^{\rm SC}}{\varepsilon_c^{\rm SC}}\right)^{-1/2} 
\left(\frac{\varepsilon}{\varepsilon_m^{\rm SC}}\right)^{-p/2},
& {\rm for}~\varepsilon_m^{\rm SC} < \varepsilon,
\end{array}
\right.
\label{eq:ssc_spectrum}
\end{equation}
where
\begin{equation}
\varepsilon_a^{\rm SC} = 4\gamma_c^2 \varepsilon_a, ~~ 
\varepsilon_c^{\rm SC} = 4\gamma_c^2 \varepsilon_c.
\end{equation}
The $\varepsilon F_{\varepsilon}$ peak energy is given by $\varepsilon_m^{\rm SC}$.
Equation (\ref{eq:SSC_num}) and (\ref{eq:SSC_fnuc}) imply that the 1st-order SSC 
radiation can explain the flux of the observed MeV emission.

The 2nd-order SSC emission is significantly suppressed by the Klein-Nishina effect.
The peak of the 2nd-order SSC spectrum is given by
\begin{equation}
\varepsilon_{\rm KN} = \Gamma_j \gamma_c m_e c^2 \frac{1}{1+z} \simeq 
38~{\rm GeV}~\Gamma_{j,3} \left(\frac{\gamma_c}{400}\right).
\end{equation}

\subsection{Spectral and temporal behavior of the UC emission}
\label{sec:aic}

Here we derive the observed spectrum of the UC emission as a function of the 
polar angle $\theta$ of the emitting region on the shell 
(see Figure~\ref{fig:geometry}). 
In the comoving frame of the jet, the seed cocoon photons are highly anisotropic.
As a result, the cocoon photons up-scattered by isotropic electrons have an anisotropic
energy distribution.
The spectrum of radiation scattered at an angle $\theta'_{\rm sc}$ relative to
the direction of the photon beam in the Thomson scattering regime is given by
\citep{aharonian81,brunetti01,wang06,fan08}
\begin{equation}
j'_{\varepsilon'}(\theta'_{\rm sc}) = \frac{3}{2}
\sigma_T (1-\cos\theta'_{\rm sc})\int d\gamma N'(\gamma) \int^1_0 dy
J'_{\varepsilon'_s}(1-2y+2y^2),
\label{eq:aic_emissivity}
\end{equation}
where $y=\varepsilon'/[2\gamma^2 \varepsilon'_s (1-\cos\theta'_{\rm sc})]$.
This is the scattered radiation emissivity in the jet comoving frame, 
$N'(\gamma)$ is the electron energy spectrum, given by equation 
(\ref{eq:electron_spectrum}),
and $J'_{\varepsilon'_s}$ is the intensity of the seed photons averaged over solid angle,
i.e., the mean intensity.
This equation can be directly derived by averaging equation (17) of \citet{brunetti01}
over the electron directions.
When this equation is averaged over $\cos\theta'_{\rm sc}$, it reduces to the 
equation for isotropic IC emission \citep{sari01}.
The term $\xi \equiv 1-\cos\theta'_{\rm sc}$ describes the anisotropy of the
spectrum, and this is due to the fact that the IC scattering is strongest
for the head-on collisions between electrons and seed photons.
This implies that the UC emission in the observer frame is stronger from the 
high-latitude region of the shell, so that its flux peak lags behind the
onset of the synchrotron and SSC emission of the same electrons, which have
isotropic energy distribution in the comoving frame of the jet.

If the direction of the photon beam is identical to the direction of the 
expansion of the scattering point of the jet, we may take $\theta'_{\rm sc} =
\theta'$ to calculate the observed flux from the scattering point with the 
polar angle $\theta'$ with respect to the jet axis measured in the jet comoving frame.
In this case
\begin{equation}
\xi \equiv 1-\cos\theta'_{\rm sc} = 1-\cos\theta' 
\approx \frac{2\Gamma_j^2 \theta^2}{1+\Gamma_j^2 \theta^2}.
\label{eq:onaxis_xi}
\end{equation}
In our case, the cocoon is viewed off-axis, and
the direction of the seed photon beam in the comoving frame is not
identical to that of the expansion (see Figure~\ref{fig:geometry}).
However, the difference is found to be negligible for our adopted parameters.
The angle between the direction of the photon beam and the expansion direction
in the lab frame is $\chi \approx (r_{\rm ph}/r_i)(\theta_j - \theta) 
\sim 10^{-4}$, since we focus on the region of $\theta < \Gamma_j^{-1}
\ll \theta_j$. 
Our case has
\begin{equation}
\xi = 1-\cos(\theta' \pm \chi') \approx \frac{2 \Gamma_j^2 (\theta \pm \chi)^2}
{(1+ \Gamma_j^2 \theta^2)(1+\Gamma_j^2 \chi^2)}.
\label{eq:offaxis_xi}
\end{equation}
Since $\Gamma_j \chi \sim 0.1$, and thus the difference between the values
of $\xi$ for the on-axis and off-axis cases is so small that we can use the 
on-axis equation (\ref{eq:onaxis_xi}). 

The mean intensity of the cocoon emission measured at $r = r_i$ in the
central engine frame is calculated as
\begin{equation}
J_{\varepsilon_s} = \frac{1}{4\pi r_i^2}\frac{d_L^2}{1+z} 
F_{\varepsilon_s}^{\rm co},
\end{equation}
and $J'_{\varepsilon'_s} = J_{\varepsilon_s}/(2\Gamma_j)$.
We perform the integration in equation (\ref{eq:aic_emissivity}) approximately
and obtain an analytical form
\begin{equation}
j'_{\varepsilon'}(\theta') \simeq \frac{3}{2} \sigma_T \xi(\theta') n' 
J'_{\varepsilon'_{\rm ph}} g(\varepsilon'),
\label{eq:cocoon_emissivity}
\end{equation}
where
\begin{equation}
J'_{\varepsilon'_{\rm ph}} = \frac{1}{4\pi r_i^2}\frac{d_L^2}{1+z} 
\frac{F_{\varepsilon_{\rm ph}}^{\rm co}}{2\Gamma_j},
\end{equation}
\begin{equation}
g(\varepsilon') = \left\{
\begin{array}{lc}
\left(\frac{\varepsilon'}{{\varepsilon_c^{\rm UC}}'}\right)^1, 
& {\rm for}~\varepsilon'<{\varepsilon_c^{\rm UC}}', \\
\left(\frac{\varepsilon'}{{\varepsilon_c^{\rm UC}}'}\right)^{-1/2}, 
& {\rm for}~{\varepsilon_c^{\rm UC}}' < \varepsilon' < 
{\varepsilon_m^{\rm UC}}', \\
\left(\frac{{\varepsilon_m^{\rm UC}}'}{{\varepsilon_c^{\rm UC}}'}\right)^{-1/2} 
\left(\frac{\varepsilon'}{{\varepsilon_m^{\rm UC}}'}\right)^{\beta},
& {\rm for}~{\varepsilon_m^{\rm UC}}'<\varepsilon'<{\varepsilon_{\rm cut}^{\rm UC}}', 
\\
\left(\frac{{\varepsilon_m^{\rm UC}}'}{{\varepsilon_c^{\rm UC}}'}\right)^{-1/2} 
\left(\frac{{\varepsilon_{\rm cut}^{\rm UC}}'}{{\varepsilon_m^{\rm UC}}'}\right)^
{\beta} \left(\frac{\varepsilon'}{{\varepsilon_{\rm cut}^{\rm UC}}'}\right)^{-p/2}, 
& {\rm for}~{\varepsilon_{\rm cut}^{\rm UC}}' < \varepsilon'.
\end{array}
\right.
\label{eq:aic_spectrum}
\end{equation}
Here we define
\begin{equation}
{\varepsilon_c^{\rm UC}}' = 2\gamma_c^2 {\varepsilon_{\rm ph}^{\rm co}}' \xi(\theta),~~
{\varepsilon_m^{\rm UC}}' = 2\gamma_m^2 {\varepsilon_{\rm ph}^{\rm co}}' \xi(\theta),~~ 
{\varepsilon_{\rm cut}^{\rm UC}}' = 2\gamma_m^2 {\varepsilon_{\rm cut}^{\rm co}}' 
\xi(\theta),
\end{equation}
and
\begin{equation}
{\varepsilon_{\rm ph}^{\rm co}}' = \frac{1+z}{2\Gamma_j} \varepsilon_{\rm ph}^{\rm co},
~~{\varepsilon_{\rm cut}^{\rm co}}' = \frac{1+z}{2\Gamma_j} 
\varepsilon_{\rm cut}^{\rm co}.
\end{equation}
The $\varepsilon F_{\varepsilon}$ peak energy of the UC emission is given by
$\varepsilon_m^{\rm UC}$ in the case of $\beta \simeq -1$ and $p \simeq 3.2$.

In order to concentrate on the time-averaged spectrum including the high-latitude
emission, we calculate the flux of the UC emission 
by neglecting the radial structure of the emitting shell for simplicity.
In other words, we consider that the emission is produced instantaneously in the 
infinitely thin shell at $r=r_i$.
This assumption gives us a model lightcurve that resembles the observed one.
Then we obtain a formula for calculating the flux of the UC emission
\begin{equation}
F_{\varepsilon}(t) = \frac{3}{2} \tau F_{\varepsilon_{\rm ph}}^{\rm co} \xi(\theta(t)) 
\frac{g(\varepsilon')}{[1+\Gamma_j^2 \theta^2(t)]^2},
\label{eq:lightcurve}
\end{equation}
where 
\begin{equation}
\varepsilon' = (1+z)\varepsilon\frac{1+\Gamma_j^2\theta^2(t)}{2\Gamma_j}
\end{equation}
\begin{equation}
\theta(t) = \sqrt{2} \left[1- \frac{c}{r_i} 
\left(\bar{t}_i - \frac{t}{1+z}\right)\right]^{1/2},
\end{equation}
and $\bar{t}_i$ is the emission time in the central engine frame
(see Appendix~\ref{sec:appendix}).

The peak energy and the peak flux of the UC emission are written as functions
of the angle parameter $q(\theta) \equiv \Gamma_j^2\theta^2$:
\begin{eqnarray}
\varepsilon_m^{\rm UC} &=& 2\gamma_m^2 \varepsilon_{\rm ph}^{\rm co} 
\frac{\xi(\theta)}{1+\Gamma_j^2 \theta^2} \nonumber \\
&\simeq& 160~{\rm MeV}~\left[\frac{4 q}{(1+q)^2}\right]
\left(\frac{\gamma_m}{400}\right)^2 
\left(\frac{\varepsilon_{\rm ph}^{\rm co}}{1~{\rm keV}}\right), 
\label{eq:co_nu} \\
\varepsilon_m^{\rm UC} F_{\varepsilon_m}^{\rm UC} &=& 3 \tau \gamma_m \gamma_c
\varepsilon_{\rm ph}^{\rm co} F_{\varepsilon_{\rm ph}}^{\rm co}  
\frac{\xi^2(\theta)}{[1+\Gamma_j^2\theta^2]^3}
\nonumber \\
&\simeq& 580~{\rm keV}~{\rm cm}^{-2}~{\rm s}^{-1}~
\left[\frac{40 q^2}{(1+q)^5}\right] \nonumber \\
&\times& \left(\frac{\tau}{4\times10^{-4}}\right) \left(\frac{\gamma_m}{400}\right)
\left(\frac{\gamma_c}{400}\right)
\left(\frac{\varepsilon_{\rm ph}^{\rm co} F_{\varepsilon_{\rm ph}}^{\rm co}}
{30~{\rm keV}{\rm cm}^{-2}{\rm s}^{-1}}\right),
\label{eq:co_fnu}
\end{eqnarray}
where the functions in the brackets $[~]$ both have values of zero at $q=0$ and 
$q=\infty$ and have peaks of $1$ at $q=1$ and $\simeq 1.4$ at $q=2/3$, respectively.
This means that the UC flux has a peak at $q \simeq 1$, or 
$\theta \simeq \Gamma_j^{-1}$, i.e.,
the peak time of the UC emission lags behind that of the SSC emission on the 
angular spreading timescale, $\Delta t_i \simeq 2$~s.
This is consistent with the observed lag of the GeV emission onset behind 
the MeV emission peak of the second pulse of GRB 080916C.
Here the values of the jet parameters $\tau = 4\times10^{-4}$ and $\gamma_m
=\gamma_c=400$ are applicable for the 1st-order SSC emission of the jet being 
consistent with the observed MeV emission component (see \S~\ref{sec:is_model}).
This indicates that the UC emission of the electrons accelerated in the internal 
shock of the jet, emitting the observed MeV emission, can naturally explain 
the observed flux in the GeV range.

\section{Constraints on model parameters}
\label{sec:fitting}

In the above two sections we have shown that the up-scattered cocoon emission model
can explain the observed properties (ii) and (iii) of GRB 080916C listed in 
\S~\ref{sec:intro}.
The high-energy lag of the second pulse (property (iii)) can be explained by
the delayed peak time of the UC emission because of
the anisotropic inverse Compton scattering.
If the constraint on the cocoon physical parameters $E_c$ and $\Gamma_c$ 
(equation \ref{eq:co_constraint1}) is satisfied, the cocoon has not released
its radiation when the first pulse is produced, and thus
the first pulse is not accompanied by the UC emission (property (ii)).
The observed high-energy spectrum of the first pulse $F_{\varepsilon} \propto 
\varepsilon^{-1.63 \pm 0.12}$ \citep{abdo09} can be assumed to be 
produced only by the 1st-order SSC emission.
This leads to an energy spectral index of the accelerated electrons 
$p \sim 3.2$ (see equation \ref{eq:ssc_spectrum}).
The first and second internal shocks may have similar electron acceleration mechanisms,
in which case we can take $p \sim 3.2$ also for the second MeV pulse.

In order to confirm whether this model can also explain the observed property (i),
i.e., the combination of the 1st-order SSC and UC emission
can reproduce the observed smooth power-law spectrum
of the second pulse, we perform numerical calculations of the
equation (\ref{eq:lightcurve}).
We approximate the spectral shape as
\begin{eqnarray}
g(\varepsilon') &=& \left(\frac{\varepsilon'}{{\varepsilon_c^{\rm UC}}'}\right) 
\left[1+\left(\frac{\varepsilon'}{{\varepsilon_c^{\rm UC}}'}\right)^s\right]^
{\frac{-3/2}{s}} \nonumber \\
&\times& \left[1+\left(\frac{\varepsilon'}{{\varepsilon_m^{\rm UC}}'}\right)^s\right]^
{\frac{\beta+(1/2)}{s}}
\left[1+\left(\frac{\varepsilon'}{{\varepsilon_{\rm cut}^{\rm UC}}'}\right)^s\right]^
{\frac{(-p/2)-\beta}{s}},
\end{eqnarray}
where we adopt the value $s=2$.
We can also similarly calculate the spectrum of the SSC emission 
(see Appendix~\ref{sec:appendix}), and combine the UC and SSC emission.

If the flux of the cocoon photospheric X-rays is given, i.e., $E_c$, $\Gamma_c$,
$r_*,$ and $\alpha$ are given, the fluxes of the UC and SSC emission of the jet
are determined by the jet parameters $L, \Gamma_j, \Delta t_i, 
\epsilon_B$, and $\epsilon_e$.
Since $\Delta t_i \sim 2$~s is roughly given by the 
observations, and this value is necessary
to explain the observed high-energy lag timescale, we have four free parameters.
On the other hand, we have four characteristic observables; the peak fluxes and 
peak photon energies of the SSC component and the UC component.
Therefore the jet parameters are expected to be constrained tightly.

Figure~\ref{fig:spectrum} shows the result of the time-averaged spectrum of
the second pulse for the cocoon parameters
\begin{equation}
E_{c,52} = 1.0,~~ \Gamma_c = 52,~~ r_{*,11} = 2.5,~~ \alpha = 0.05,
\label{eq:cocoon_parameter}
\end{equation}
and $\beta = -1.2$.
These values satisfy the constraints on the cocoon parameters, equations
(\ref{eq:co_constraint1}), (\ref{eq:co_constraint2}), and (\ref{eq:co_constraint3}).
The {\it dashed line} represents the 1st-order SSC component plus the 2nd-order SSC
component without taking account of Klein-Nishina effect, the {\it dot-dashed line}
represents the cocoon X-ray emission, and the {\it dotted line} represents the 
UC component.
The {\it thick solid line} is the combination of
all the components taking into account the Klein-Nishina effect.
The {\it thin solid line} and the {\it dot-short-dashed lines} represent the Band 
model spectrum fitted to the observed data and the 95\% confidence errors, 
respectively (from the {\it Fermi} LAT/GBM analysis group, private communication).
This figure shows that our model is roughly consistent with the observed spectrum
at $\varepsilon \gtrsim 1$~MeV.
Since the observed number flux of the second pulse at $\sim 1$~GeV is 
$\sim 1~{\rm counts}/{\rm s}$, the detection limit for $\varepsilon > 3$~GeV
is roughly given by $\varepsilon F_{\varepsilon,{\rm lim}} \sim 10^3~{\rm keV}~
{\rm cm}^{-2}~{\rm s}^{-1} (\varepsilon/3~{\rm GeV})$
\citep[The effective area of the LAT detector is approximately constant for
$1~{\rm GeV} \lesssim \varepsilon \lesssim 300~{\rm GeV}$,][]{atwood09}.
Thus the small bump at $\sim 30$~GeV cannot be detected.
This is consistent with the non-detection of photons at $> 3$~GeV.
The adopted values of the jet parameters are
\begin{equation}
L_{55} = 1.1,~~ \Gamma_{j,3} = 0.93,~~ \Delta t_i = 2.3~{\rm s},~~
\epsilon_B = 10^{-5},~~\epsilon_e = 0.4,
\label{eq:jet_parameter}
\end{equation}
and $p = 3.2$.
The corresponding values of the optical depth for electron scattering and 
the characteristic electron Lorentz factors are
\begin{equation}
\tau = 3.5 \times 10^{-4},~~ \gamma_m = 400, ~~ \gamma_c = 390.
\end{equation}
Figure~\ref{fig:lightcurve1} shows the results of the multi-band lightcurves
for the same parameters.
Each lightcurve is normalized to a peak flux of unity.
This clearly displays the lag of the high-energy emission onset.
We also plot the multi-band lightcurves averaged in 0.5~s time bins in
Figure~\ref{fig:lightcurve2}, which corresponds to the lightcurves measured
by the GBM and LAT of {\it Fermi} satellite \citep[Fig 1 of][]{abdo09}.
The observed peak number flux of the GeV photons is only 
$\sim 1~{\rm counts}/{\rm bin}$,
so that the rising part of the GeV emission at $t \leq 5.5$~s could not be detected.
This is consistent with the onset of the GeV emission being delayed behind 
that of the MeV emission.

\section{Discussion}
\label{sec:discussion}

\subsection{An additional emission component in the sub-MeV range?}
\label{sec:discussion1}

The 1st-order SSC emission and UC emission of the internal shock electrons of 
the jet can explain the main spectral peak (at $\varepsilon \sim 1$~MeV) and 
the delayed high-energy part (at $\varepsilon \gtrsim 100$~MeV), respectively, 
of the second pulse of GRB 080916C (see Figure~\ref{fig:spectrum}).
However, we should note a caveat for this model.
The 1st-order SSC emission has a steep low-energy slope, i.e., 
$\varepsilon F_{\varepsilon} \propto \varepsilon^2$ at $\varepsilon < 1$~MeV
because of strong synchrotron self-absorption effect 
(see equation \ref{eq:self_absorption}), which is not consistent with the
observed low-energy spectrum $\varepsilon F_{\varepsilon} \propto 
\varepsilon^{0.98 \pm 0.02}$ \citep{abdo09}.
On the other hand, the 2nd-order SSC emission has the same steep low-energy slope, 
which is consistent with the non-detection of the spectral bump at 
$\varepsilon \gtrsim 1$~GeV.
If there was no self-aborption effect, the 2nd-order SSC emission, even with
the Klein-Nishina effect, would be so bright as to be detected \citep[see also][]{wang09}.
In order to reproduce the observed sub-MeV spectrum, we require an additional
emission component, with careful consideration of the physical conditions of the
emitting region.
An additional sub-MeV emission produced by some process {\it near the emission
region of the main MeV emission of the second pulse} would be up-scattered by
the electrons emitting the MeV emission, like the SSC process without the 
self-absorption, and then the up-scattered flux would easily violate the 
observational upper limits.

Here we propose, for illustrative purposes, an additional sub-MeV emission 
whose up-scattered emission does not violate the observational limits.
One can argue that additional sub-MeV emission may be produced in another internal 
shock (which we can call a ``third internal shock") occurring far in front of the 
second internal shock shell at the same observed time. 
A fraction of the emission from the third internal shock propagates backward
and is up-scattered by the second shell electrons into the high-energy range,
but in this case, the electrons have cooled by the adiabatic expansion and then
the up-scattered flux is expected to be dimmer than the case mentioned above.

Figure~\ref{fig:spectrumff} demonstrates an example of such an additional sub-MeV
emission.
The {\it dashed line} is the same as the thick solid line in Figure~\ref{fig:spectrum},
i.e., the time-averaged spectrum of the combination of the SSC and UC emission
of the second internal shock and the cocoon X-ray emission.
The {\it dotted line} represents the synchrotron emission of the third internal shock.
Here we assume that this third, less energetic internal shock accelerates some
fraction $f$ of electrons with a power-law energy distribution
\citep{bykov96,eichler05,toma08,bosnjak09}
\footnote{
We find it difficult to obtain an additional sub-MeV component by the SSC
mechanism similar to the main MeV component.
A sub-MeV component should be less energetic than the main MeV component, so that
it should be $\gamma_m < \gamma_c$.
In this case the flux at $\varepsilon_m^{\rm SC}$ should make a sub-MeV peak.
Then $\varepsilon_m^{\rm SC}$ should be about one order of magnitude smaller than and
$F_{\varepsilon_m}^{\rm SC}$ should be comparable with those for the main MeV 
component, and $\varepsilon_a/\varepsilon_m$ should be smaller than unity.
However, such values may not be obtained by equation (\ref{eq:SSC_num}), 
(\ref{eq:SSC_fnuc}) with $F_{\varepsilon_c}^{\rm SC} \to F_{\varepsilon_m}^{\rm SC}$ 
for $\gamma_m < \gamma_c$, and (\ref{eq:self_absorption}) with 
$\varepsilon_c \to \varepsilon_m$ and $\gamma_c \to \gamma_m$ for any reasonable set
of the jet parameter values 
$\{L, \Gamma_j, \Delta t_i, \theta_p, \epsilon_B, \epsilon_e \}$.
}.
The adopted parameters are
\begin{eqnarray}
L_{55} &=& 0.23, ~~\Gamma_{j,3} = 1.8, ~~\Delta t_i = 2.3~{\rm s}, \nonumber \\
\epsilon_B &=& 1.3\times10^{-3}, ~~\epsilon_e = 0.35, ~~f=3\times10^{-3}.
\label{eq:jet_parameter_ad}
\end{eqnarray}
The corresponding values of the scattering optical depth for the accelerated electrons
and the characteristic electron Lorentz factors are
\begin{equation}
\tau_{ac} = 8.2\times10^{-9}, ~~ \gamma_m = 1.2\times10^5, ~~\gamma_c = 5.6\times10^4. 
\end{equation}
The SSC emission of the third internal shock electrons are significantly suppressed
by the Klein-Nishina effect.
The cocoon X-rays and the optical synchrotron emission of the second internal shock
are up-scattered by the third internal shock efficiently, whose fluxes have been
taken into account for estimating $\gamma_c$.
The former has the peak flux $\varepsilon F_{\varepsilon} \sim 200~{\rm keV}~
{\rm cm}^2~{\rm s}^{-1}$ at $\varepsilon \sim 10$~TeV.
This may be absorbed by the extragalactic background radiation.
The latter is shown by the {\it dot-dashed line} in Figure~\ref{fig:spectrumff},
which may arrive at the earth along with the second pulse in the high-energy range
but be so dim as not to be detected by the LAT detector.
The emission radius of the third shell is estimated by
\begin{equation}
r_{ia} \simeq 8.4\times10^{16}~{\rm cm}.
\end{equation}
Thus the radius where the backward photons from the third shell is up-scattered
by the second shell is $\simeq (r_i + r_{ia})/2 \simeq 5.3 \times 10^{16}~{\rm cm}
\equiv R_{\rm ex} r_i$ where $R_{\rm ex} \simeq 2.4$.
At this radius, the column density of the second shell is $R_{\rm ex}^2$ times
smaller and the characteristic electrons Lorentz factors are $R_{\rm ex}$ times
smaller because of the adiabatic expansion, so that the optical depth for the 
up-scattering is $R_{\rm ex}^4$ times smaller than that at $r=r_i$.
The synchrotron emission from the third shell, up-scattered by the expanded
second shell, is calculated by using equation (\ref{eq:cocoon_emissivity}) with
$\xi = 1-\cos(\pi-\theta')$, and the result is shown by the {\it thick dashed line}
in Figure~\ref{fig:spectrumff}.
This emission arrives at the earth $(R_{\rm ex}-1) \Delta t_i \simeq 3.2$~s after
the onset of the second pulse, and may not be detected by the LAT detector.
The combined spectrum of the second internal shock emission ({\it dashed line}),
the synchrotron emission of the third internal shock ({\it dotted line}), and
the synchrotron emission of the second internal shock up-scattered by the third
internal shock electrons ({\it dot-dashed line}) is shown by the {\it thick solid 
line} and it may be observed as the second pulse.
This is consistent with the observed spectrum of the second pulse.

\subsection{Internal shock radius vs deceleration radius}

The jet material is finally decelerated by the interaction with the circumburst
medium, into which the external shock propagates.
Thus the internal shocks of the jet must occur below the deceleration radius of 
the first internal shock shell.
The first shell is estimated to have the isotropic-equivalent kinetic energy
$E \sim L \Delta t_i/(1+z) \simeq 4.7\times10^{54}$~erg and the bulk Lorentz
factor $\Gamma_j \sim 10^3$ after the production of the first MeV pulse.
The deceleration radius is then $r_{\rm dec} \simeq 
[17 E/(16\pi \Gamma_j^2 n m_p c^2)]^{1/3} \sim 10^{17}~{\rm cm}~n^{-1/3}$.
This would be larger than the internal shock radius $r_i$ or $r_{ia}$ if
$n \lesssim 1~{\rm cm}^{-3}$.
This is in a reasonable range of typical values of the GRB circumburst medium
density \citep{panaitescu02,yost03}.
The deceleration time is estimated to be
$t_{\rm dec} \simeq r_{\rm dec}(1+z)/(2c\Gamma_j^2) \sim 9~ n^{-1/3}$~s.
After the UC emission ceases (i.e., $t > \Delta t_c + \Delta t_d \sim 15$~s) 
the high-energy emission may be due to the external shock \citep[see][]{kumar09b}.

\subsection{Photospheric emission of the jet}
\label{sec:thermal_jet}

We have assumed that the jet is dominated by kinetic energy of baryons instead of
magnetic energy. 
\citet{zhang09} have argued that the non-detection of the photospheric thermal emission 
{\it of the jet} for GRB 080916C may exclude a possibility of a baryon-dominated 
jet and suggest that a Poynting-flux-dominated jet is preferred
\citep[see also][]{daigne02}.
They show that if the jet is dominated by thermal energy at $r = c \Delta t_i/(1+z)
\simeq 2.8 \times 10^{9}~(\Delta t_i/0.5~{\rm s})$~cm, 
the photospheric thermal emission of the jet is so bright that it can be detected, 
which is inconsistent with the observation.
However, it is more reasonable that the jet has already accelerated up to 
a $\Gamma_j \approx 10^3$ before reaching that radius, through the usual
adiabatic expansion starting from the central region of the progenitor star 
($r_e \sim 10^6$~cm. This value corresponds to the gravitational radius of a 
black hole with a mass $M = 3M_{\odot}$).
Thus the jet is dominated by kinetic energy at that radius, and 
the thermal emission of the jet may be much weaker than the estimate by
\citet{zhang09}.

We assume that the jet is dominated by thermal energy at $r = r_e$, and it
is accelerated by converting the thermal energy into kinetic energy.
We can derive the flux of the residual thermal emission of the jet at the photosphere
according to the standard fireball model, similar to that of the cocoon 
(see \S~\ref{sec:cocoon_dynamics}).
Let the initial bulk Lorentz factor of the jet at $r = r_e$ be 
$\Gamma_e \geq 1$. 
(It may be possible that a fraction of the black hole rotational energy is carried 
into the kinetic energy of the jet by the magnetic field.)
Then the bulk Lorentz factor of the jet saturates at 
$r_{sj} = r_e \Gamma_j / \Gamma_e \simeq 10^9~{\rm cm}~r_{e,6} \Gamma_{j,3}
\Gamma_e^{-1}$.
The comoving number density of the jet shell $n' = L/(4\pi r^2 m_p c^3 \Gamma_j^2)$
and the Thomson optical depth $\tau = \sigma_T n' r/\Gamma_j = 1$ define the 
photospheric radius 
\begin{equation}
r_{\rm phj} = \frac{\sigma_T L}{4\pi m_p c^3 \Gamma_j^3}
\simeq 1.2 \times 10^{13}~{\rm cm}~L_{55} \Gamma_{j,3}^{-3}.
\end{equation}
The comoving temperature of the jet shell evolves from $T'_e \simeq 
[L/(4\pi r_e^2 c \Gamma_e^2 a)]^{1/4}$ as $\propto r^{-1}$\ 
at $r_e < r < r_{sj}$ and as $\propto r^{-2/3}$ at $r_{sj} < r < r_{\rm phj}$.
Then we obtain the peak energy and the peak flux of the photospheric emission of 
the jet shell
\begin{eqnarray}
\varepsilon_{\rm ph}^{\rm jet} &\simeq& 43~{\rm keV}~L_{55}^{-5/12} r_{e,6}^{1/6}
\Gamma_e^{-1/6} \Gamma_{j,3}^{8/3}, \\
\varepsilon_{\rm ph}^{\rm jet} F_{\varepsilon_{\rm ph}}^{\rm jet} &\simeq&
460~{\rm keV}~{\rm cm}^{-2}~{\rm s}^{-1}~L_{55}^{1/3} r_{e,6}^{2/3} \Gamma_e^{-2/3}
\Gamma_{j,3}^{8/3}.
\end{eqnarray}
(The time-averaged flux is smaller by about a factor of 4.)
This means that the photospheric thermal emission of the jet is marginally hidden
by the prompt MeV emission (see Figure~\ref{fig:spectrumff}).

The above analysis does not take account of possible effects from the stellar matter.
The front of the jet is dissipated by the interaction with the stellar matter,
and the shocked jet matter and the shocked stellar matter escape sideways to form
the cocoon. The shocked jet matter in front of the clean jet at the break-out
will make an X-ray flash. The brightness of the X-ray flash highly depends on the density 
profile of the star \citep{waxman03}. We do not discuss this in further detail.

If re-conversion of jet kinetic energy into thermal energy due to interaction 
of the sides of the jet with the high-pressure cocoon 
is strong, the residual thermal emission may be brighter than the above estimate.
The ratio of thermal energy to kinetic energy of
the clean, adiabatically expanding jet shell at the stellar radius is 
$\simeq (r_*/r_{sj})^{-2/3} \sim 0.05$.
Thus the re-conversion fraction at $r \leq r_*$ is required to be smaller than 5\%
to suppress the thermal flux down to the above estimate.
However, many numerical simulations with various initial conditions of the jet and the star
show that the re-conversion fractions can be much higher than 5\%
while the jet propagates in the star \citep[e.g.,][]{wzhang03,mizuta06,morsony07,mizuta09}
and even well outside the star \citep{lazzati09}.
This implies that the photospheric thermal emission of the jet for GRB 080916C may not 
be hidden by the SSC emission from the internal shock, being inconsistent with the 
observation.

It could be possible that the prompt MeV emission of GRB 080916C is mainly the 
photospheric emission of the jet \citep{meszaros00,lazzati09}. 
In this case also, the delayed high-energy emission could be due to the up-scattering 
of the cocoon emission off the non-thermal electrons produced in internal shocks, 
where the SSC and synchrotron emission from the internal shocks
are dimmer than the photospheric emission in the MeV energy range. This scenario
requires different modeling from that in this paper.

Alternatively, as discussed by \citet{zhang09}, it is possible that the jet
is initially dominated by magnetic energy and the baryons in the jet are accelerated
by converting the magnetic energy into kinetic energy, so that the jet is 
dominated by the baryon kinetic energy at the internal shock radius
\citep[e.g.,][]{vlahakis03,komissarov09,lyubarsky09}. 
The cocoon will be still produced by the shocked jet and shocked stellar matter, but
a large fraction of the cocoon energy will be the magnetic energy. However, the magnetic
fields could be tangled and dissipated through escaping from the jet into the cocoon and 
interacting with the shocked stellar matter and the ambient stellar envelope, so that 
the magnetic energy could become in equipartition to thermal (radiation) energy. 
The dynamics of such a cocoon after escaping from the star is similar to that 
in the non-magnetized jet case.
This possibility is speculative, and magnetohydrodynamic simulations of the jet
propagation in a star will be useful to address it.
If this is true, our model based on the assumption that the MeV and high-energy emission
components are produced by the 1st-order SSC radiation from the internal shocks and the
up-scattered cocoon radiation, respectively, will be applicable.
It is an open question how bright the photospheric thermal emission from the
magnetized jet can be.

\section{Implications for other long and short GRBs}
\label{sec:other}
 
We have focused on the up-scattered cocoon emission for explaining the delayed 
high-energy emission of GRB 080916C.
Since the delayed high-energy emission is observed for other bursts,  
it is useful to discuss whether the up-scattered cocoon emission may be
important for other bursts.

Our model involves many parameters; the cocoon parameters $E_c$, $\Gamma_c$,
$r_*$, and $\alpha$, and the jet parameters $L$, $\theta_j$, $\Gamma_j$,
$\Delta t_i$, $\epsilon_B$, and $\epsilon_e$. The cocoon parameters depend 
significantly on the luminosity and the initial opening angle of the jet and the 
density profile of the progenitor star \citep[e.g.,][]{meszaros01,morsony07},
so that their values for general bursts are unclear. Here, for simplicity,
we fix the values of the cocoon parameters as those adopted above for GRB 080916C.

For further simplicity, we only consider the bursts for which the prompt 
emission in the soft $\gamma$-ray band could be produced by the 1st-order
SSC radiation of the electrons. Indeed, it seems difficult to explain the 
prompt emission of {\it all} GRBs in the simple SSC model 
\citep{piran09,kumar09}.
The jet parameters for typical long GRBs are inferred from various observations;
$L \simeq 10^{53}~{\rm erg}~{\rm s}^{-1}$, $\theta_j \simeq 0.1$ \citep{ghirlanda04}, 
and $\Gamma_j \simeq 300$ \citep[e.g.,][]{lithwick01,molinari07}. 
The other parameters with values similar to those adopted for GRB 080916C
in this paper, i.e., $\Delta t_i/(1+z) \simeq 0.4$~s, $\epsilon_B \simeq 10^{-5}$,
and $\epsilon_e \simeq 0.2$ may provide $\gamma_c \sim \gamma_m$ and
account for the spectral peak energy 
$\varepsilon_m^{\rm SC} \sim 200$~keV and spectral peak flux
$\varepsilon_m^{\rm SC} F_{\varepsilon_m}^{\rm SC} \sim 
100~{\rm keV}~{\rm cm}^{-2}~{\rm s}^{-1}$ of typical long GRBs with redshifts 
$z \simeq 2$ in the SSC model (see \S~\ref{sec:is_model}). 

In this case we have the value of $\tau \gamma_m \gamma_c$ similar to that 
for GRB 080916C, so that the up-scattered cocoon emission could be important
for those bursts (see equation \ref{eq:co_fnu}), although we require to 
study some kinematic effects carefully.
Now we have
\begin{equation}
\Gamma_c \theta_j \simeq 5 > 1,
\end{equation}
so that the de-beaming effect of the relativistic motion of the cocoon
shell for the seed photons of the UC emission is significant.
However, this effect may be cancelled out by the beaming effect of the 
relativistic motion of the jet.
The internal shock radius is $r_i \simeq 2 \times 10^{15}$~cm, which is 
still much larger than the photospheric radius of the cocoon $r_{\rm ph}
\simeq 2 \times 10^{14}$~cm. 
Then we have the angle between the direction of the photon beam and the
expansion direction of the scattering point of the jet in the lab frame 
$\chi \approx (r_{\rm ph}/r_i)(\theta_j-\theta) \sim 10^{-2}$ 
(see Figure~\ref{fig:geometry}) and 
\begin{equation}
\Gamma_j \chi \sim 3 > 1.
\end{equation}
Since $\Gamma_c \theta_j \sim \Gamma_j \chi$, the de-beaming and beaming 
effects may compensate each other, so that the seed radiation can have 
the flux and spectrum similar to those for GRB 080916C. 
Therefore it is quite possible that the up-scattered cocoon emission 
contributes to typical long GRBs and they account for the delayed
high-energy emission.

The condition $\Gamma_j \chi > 1$ means that $\chi > \theta$ (since
we focus on the region of $\theta \sim \Gamma_j^{-1}$) and 
$\xi \approx 2/(1+\Gamma_j^2 \theta^2)$. Then these bursts will not
have the high-energy lags, i.e., the property (iii) (see equation
\ref{eq:co_fnu}). Since the seed cocoon photons come from the front
in the jet frame, the UC emission in the observer frame is stronger
from the line of sight.

For bursts with smaller $\Delta t_i/(1+z)$ so that $r_i \lesssim r_{\rm ph}$,
the delay timescale of the up-scattered cocoon emission behind the onset 
of the internal shock emission should be on the order of
$(1+z)r_{\rm ph} \theta_j/c \sim 10^3$~s,
and thus in this case, the up-scattering effect may be important only 
if the duration of the prompt emission is longer than $\sim 10^3$~s.

Some short GRBs might originate from the collapses of the massive stars
\citep{janiuk08,zhang09b,toma05}, and thus they could have the delayed
UC emission.
Even if other short GRBs are produced by the compact star mergers,
it might be possible that the jet is accompanied by the delayed disc wind
\citep{metzger08} and that the emission from the disc wind is 
up-scattered by the electrons accelerated in the jet.
For either progenitor models, the delayed high-energy emission 
associated with short GRBs, if any, would provide an interesting
tool to approach their origins.

\section{Summary and further implications}
\label{sec:summary}

\begin{figure}
\plotone{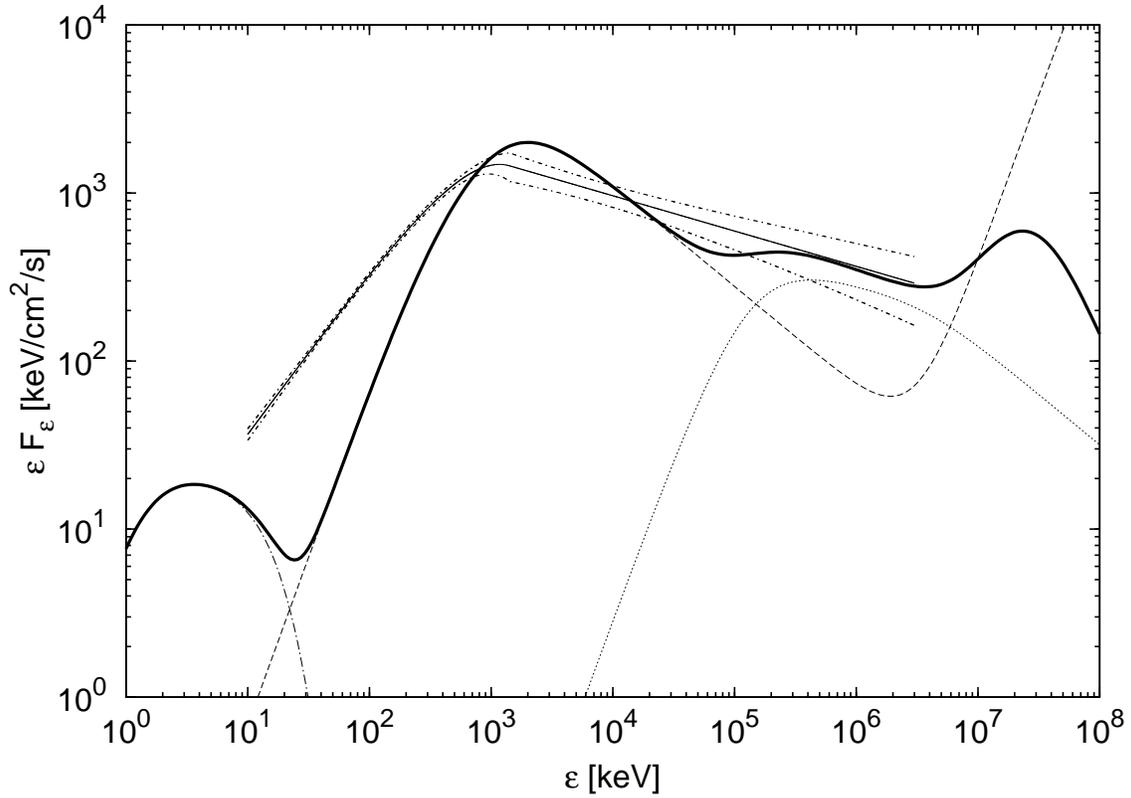}
\caption{
Time-averaged spectrum of the second pulse calculated 
in the up-scattered cocoon emission model.
The 1st-order SSC component plus 2nd-order SSC component without including the
Klein-Nishina effect ({\it dashed line}), the cocoon photospheric emission
({\it dot-dashed line}), and the up-scattered cocoon (UC) emission ({\it dotted
line}) are shown.
The {\it thick solid line} represents the combination of these, taking account of 
the Klein-Nishina effect, which is roughly consistent, at $\varepsilon \gtrsim 1$~MeV,
with the Band model spectrum ({\it thin solid line}) with 95\% confidence errors 
({\it dot-short-dashed lines}) (from the LAT/GBM group of {\it Fermi}).
The bump at $\sim 30$~GeV is so dim as not to be detected.
The adopted parameters are listed in equation (\ref{eq:cocoon_parameter}) and 
(\ref{eq:jet_parameter}).
}
\label{fig:spectrum}
\end{figure}

\begin{figure}
\plotone{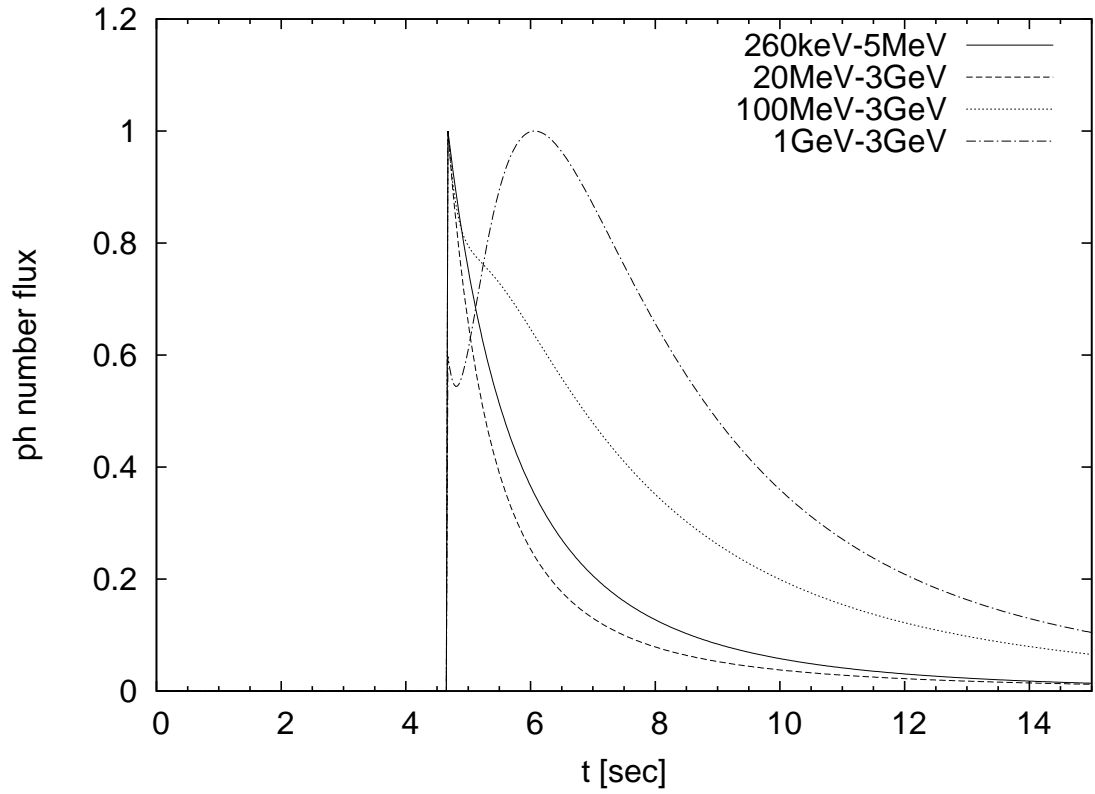}
\caption{
Photon number fluxes in several frequency ranges calculated in the up-scattered
cocoon emission model.
Time resolution is set to be 25~ms.
Each lightcurve is normalized to a peak flux of unity.
The peak of the GeV lightcurve is delayed behind that of the MeV lightcurve.
}
\label{fig:lightcurve1}
\end{figure}

The {\it Fermi} satellite provides an unparalleled broad energy coverage with
good temporal resolution, and it is accumulating a steady stream of new data on
the high-energy emission of GRBs.
GRB 080916C, detected by {\it Fermi}, showed unexpected properties in the high-energy
range, i.e., $> 100$~MeV, which are listed in \S~\ref{sec:intro}.
These include the new results on prompt emission; (i) the time-resolved high-energy
spectra are described by smooth power-laws up to a photon with energy 
$\approx 13.2$~GeV, (ii) the high-energy emission is not detected with the first
MeV pulse (i.e., the first pulse in the $\lesssim 1$~MeV range) and begins to be
detected along with the second MeV pulse, 
(iii) the second pulse has the later peaks in the higher energy bands,
and (iv) the high-energy emission lasts at least $1400$~s, much longer than the 
duration of the MeV emission.
We have discussed a model in which the prompt emission spectrum consists of an
MeV component produced by the SSC emission of electrons accelerated in 
internal shocks in the jet and the high-energy component produced by up-scattering 
of the cocoon X-rays off the same electrons, and we have shown that this model can explain 
the above three observed properties (i), (ii), and (iii).
The expanding cocoon may become optically thin some time later than the first internal
shock of the jet (equation \ref{eq:delta_t_c}), so that the first MeV pulse may not
be associated with the up-scattered cocoon (UC) emission while the second MeV pulse
may be associated with it (property (ii), see \S~\ref{sec:cocoon_dynamics}).
The UC emission has an anisotropic energy distribution in the comoving frame of the
jet so that the observed UC emission is stronger from the higher-latitude region
of the shell.
This results in the lag of the flux peak of the UC emission behind the MeV emission
onset on the angular spreading timescale (property(iii), see 
\S~\ref{sec:emission_formalism}, 
Figure~\ref{fig:lightcurve1}, and \ref{fig:lightcurve2}).
Figure~\ref{fig:spectrum} shows that the combination of the SSC and UC emission
can reproduce the observed high-energy spectral data 
(property (i), see \S~\ref{sec:fitting}).
The UC emission is short-lived and may not account for the whole high-energy
emission which lasts longer than the MeV emission (property (iv)).
It is natural that the high-energy emission in the later times is related to the 
afterglow. This has been shown by \citet{kumar09b}. However, the early portion of
the high-energy emission should be the UC emission, because the external shock
cannot reproduce the observed steep rise of the flux.

The observed flux and width of the second pulse of this burst indicate
that the isotropic-equivalent luminosity of the jet 
$L \simeq 10^{55}~{\rm erg}~{\rm s}^{-1}$ 
and the angular spreading timescale of the pulse $\Delta t_i \simeq 2$~s.
The absence of a $\gamma\gamma$ absorption cutoff in the spectra leads to
a jet bulk Lorentz factor of $\Gamma_j \simeq 10^3$ \citep{abdo09}
\citep[see also][]{aoi09}.
(A larger $\Gamma_j$ would make the internal shock radius larger than the 
deceleration radius of the shell.)
Then in the internal shock model with a mean random Lorentz factor of the shocked
protons $\theta_p \lesssim 8$ (and all of the electrons being accelerated, i.e., 
$f = 1$), the spectral peak flux at the MeV energy cannot be produced by synchrotron
emission (see equation \ref{eq:syn_num}), but can be produced
by the 1st-order SSC emission of the electrons.
The observed peak energy and the peak flux of the MeV emission constrains the 
1st-order SSC emission mechanism requiring the microphysical parameters as 
$\epsilon_B \simeq 10^{-5} \ll \epsilon_e \simeq 0.4$, or 
$\gamma_m \sim \gamma_c \sim 400$.
These parameters naturally make the UC emission of the same electrons compatible 
with the observed flux at the GeV energy, and $\Delta t_i \simeq 2$~s is 
consistent with the observed lag of the GeV emission behind the onset of the 
MeV emission.

In our model, we expect this burst to have had bright synchrotron emission 
in the optical band, like the "naked-eye" GRB 080319B \citep{racusin08}.
The 1st-order SSC emission peaking at the MeV energy has a steep slope in the
lower energy range because of a strong synchrotron self absorption effect
(see Figure~\ref{fig:spectrum}).
This effect, along with the Klein-Nishina effect, suppresses the 2nd-order SSC
emission so it is not detected in the $\varepsilon > 1$~GeV range.
On the other hand, an additional emission component is required to reproduce the 
observed spectrum in the sub-MeV energy range.
We propose that another internal shock with the smaller luminosity 
$L \simeq 2\times10^{54}~{\rm erg}~{\rm s}^{-1}$ and a fraction of electrons 
accelerated $f \simeq 3\times10^{-3}$ may produce the sub-MeV emission 
(see \S~\ref{sec:discussion1} and Figure~\ref{fig:spectrumff}).
This interpretation is somewhat ad hoc but plausible, and we argue that
it is compatible with the {\it Fermi} observation.

Many numerical simulations suggest that the hydrodynamical jet will store a
large fraction of thermal energy because of the interaction of the jet
with the high-pressure cocoon \citep{morsony07,lazzati09}. Then the jet will emit a
bright photospheric thermal radiation, which is not observed for GRB 080916C
\citep{zhang09}. It could be possible that the prompt MeV emission
is mainly the photospheric emission of the jet (instead of the 1st-order SSC 
radiation from the internal shocks), while the high-energy emission component 
is still due to the up-scattered cocoon emission from the internal shocks.
Alternatively, the jet of GRB 080916C may be initially dominated by magnetic energy.
It is not clear yet whether a large fraction of thermal energy will be stored
in the magnetized jet while it propagates in a star. The jet could become 
dominated by baryon kinetic energy at large radius, and the cocoon dynamics 
could be similar to that in the non-magnetized jet case, so that 
our model in this paper would be entirely applicable (see \S~\ref{sec:thermal_jet}). 

GRB 080916C is the only burst for which the observational data is
extensive enough and the analysis is fairly complete at the moment, so that
we have focused on this GRB to discuss our model in detail. 
Since our model involves much information from the observational data
and many parameters, it is not easy to perform a more general study 
of the UC emission at this stage.
However our model is generic, and we have briefly shown that it could apply
to other {\it Fermi}-LAT GRBs with typical parameters 
$L \simeq 10^{53}~{\rm erg}~{\rm s}^{-1}$, $\theta_j \simeq 0.1$,
and $\Gamma_j \simeq 300$ for which the prompt emission in the soft
$\gamma$-ray range is produced by the 1st-order SSC radiation of
electrons (see \S~\ref{sec:other}).
We may do a comprehensive study of our model
when more {\it Fermi}-LAT GRBs are analyzed.
The up-scattering effect could apply to short GRBs as well.
Investigating the seed photons for the delayed high-energy emission
of short GRBs would be an interesting approach to their origins.

A simple prediction of our model is that prompt emission spectra of at least
some long GRBs would have an excess above the Band spectrum around $\sim 1$~keV 
due to the cocoon photospheric emission, and this excess should have a different 
temporal behavior from that of the MeV emission.
Since the {\it Swift} XRT detector can observe the GRB prompt emission in the 
$0.3 - 10$~keV range at least 100~s after the triggers, the cocoon X-ray emission
could in principle be observed if its onset times are $t_{\rm on} > 100$~s. 
In this case, we could examine the up-scattering effect of the cocoon emission if
$t_{\rm on}$ is smaller than the burst duration $T$, while the cocoon emission may
contribute to the early X-ray afterglow if $t_{\rm on} \gtrsim T$ \citep{peer06}.
For long GRBs with $t_{\rm on} < 100$~s, the up-scattering effect could be tested
by future satellite, e.g., {\it EXIST}
\footnote{
http://exist.gsfc.nasa.gov
}, which is designed to have a soft X-ray detector and a good triggering capability.

If our model is correct, we can constrain the parameter range for which
hadronic effects are important on the high-energy emission of GRBs, and 
we can also constrain the models of high-energy cosmic ray acceleration.
Also, the delay time of the onset of the high-energy emission, $t_{\rm on}$,
is directly linked to the optical-thinning time of the expanding cocoon, which
constrains the physical parameters of the progenitor star and the cocoon material
of GRBs.
For GRB 080916C, the stellar radius $r_*$ and the total energy $E_c$ and mass $M_c$
of the cocoon are constrained to be 
$0.3 < E_{c.52}^{-2} (M_c/10^{-4}M_{\odot})^{2.5} \lesssim 0.9$ and 
$r_{*,11} \gtrsim 0.8~E_{c,52}^5 (M_c/10^{-4}M_{\odot})^{-3}$
(see equations \ref{eq:co_constraint1} and \ref{eq:co_constraint3}). 
The cocoon energy and the cocoon mass come from the jet energy 
released within the star and the stellar mass swept by the jet, respectively.
These constraints therefore provide potential tools for investigating the 
structure of the progenitor star just before the explosion, as well as the physical 
conditions of the jet propagating inside the stellar envelope through either 
analytical \citep[e.g.,][]{meszaros01,matzner03,toma07} or numerical
\citep[e.g.,][]{wzhang03,mizuta06,morsony07} approaches.

\acknowledgements
We thank K.~Murase and the {\it Fermi} LAT/GBM group members 
(especially F.~Piron and V.~Connaughton, who provided the Band model spectrum
fitted to the observed data with errors) for useful discussions.
We are grateful to the referee, D.~Lazzati, for several important comments,
and to Y.~Z.~Fan, S.~Inoue, K.~Ioka, T.~Nakamura, R.~Yamazaki, and B.~Zhang
for useful discussions after the submission of this paper.
We acknowledge NASA NNX09AT72G, NASA NNX08AL40G, and NSF PHY-0757155 for partial support.
XFW was supported by the National Natural Science Foundation of China
(grants 10503012, 10621303, and 10633040), National Basic Research Program
of China (973 Program 2009CB824800), and the Special Foundation for the 
Authors of National Excellent Doctorial Dissertations of P. R. China by
Chinese Academy of Sicences.


\clearpage

\begin{figure}
\epsscale{0.8}
\plotone{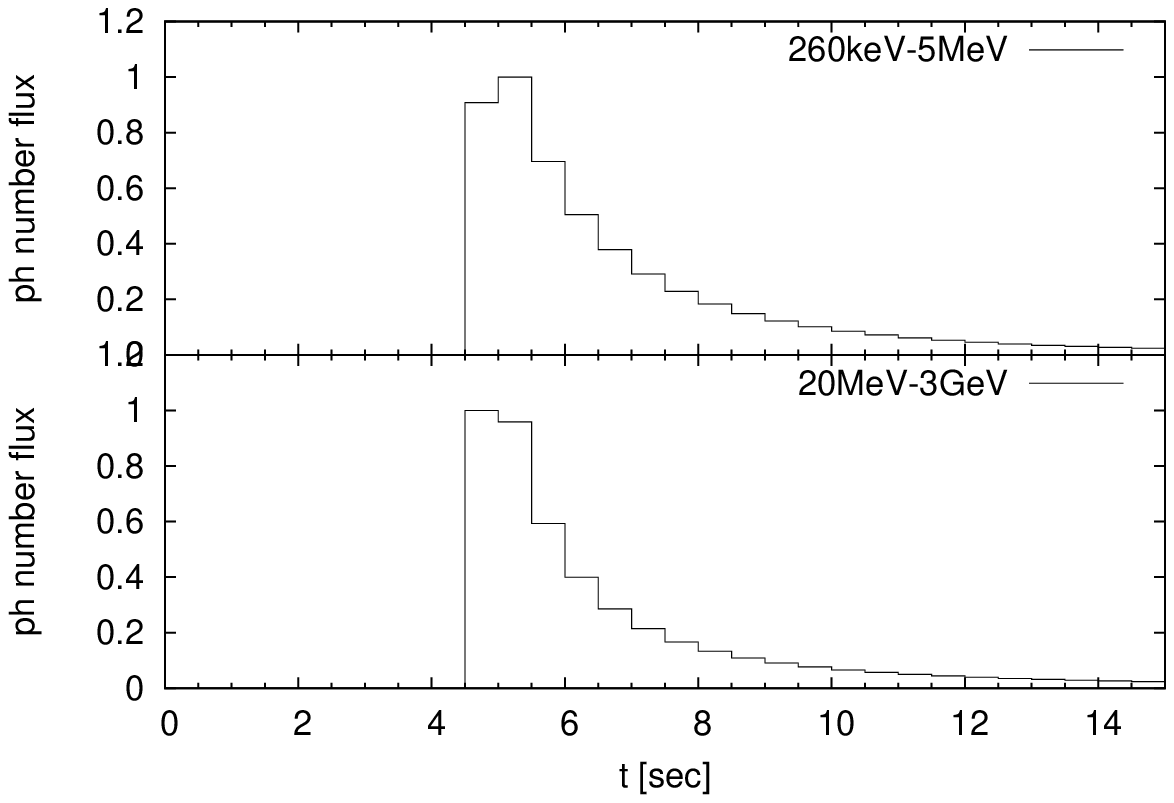}
\end{figure}
\begin{figure}
\epsscale{0.8}
\plotone{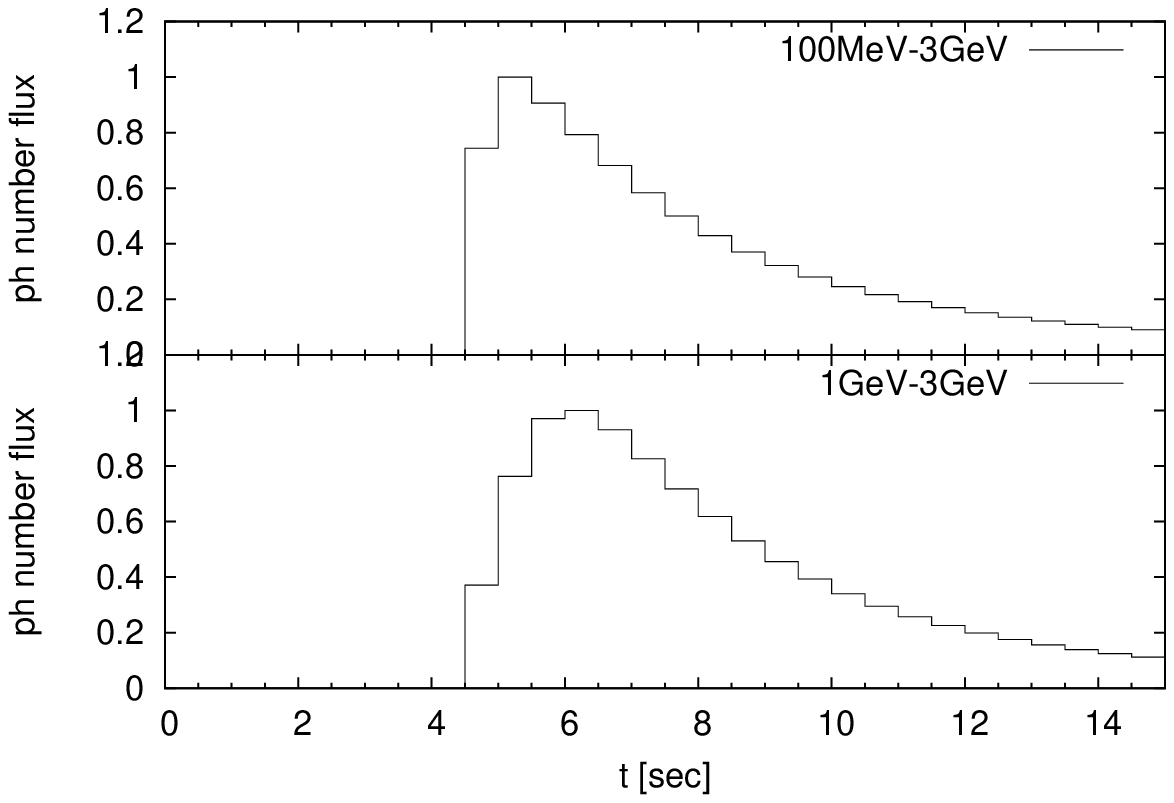}
\caption{
Same as Figure~\ref{fig:lightcurve1}, but integrated in 0.5~s time bins.
The observed number flux of the GeV photons is only $\sim 1$~counts/bin 
\citep{abdo09}, so that only the delayed peak at 
$5.5~{\rm s} \lesssim t \lesssim 7~{\rm s}$ could be detected.
}
\label{fig:lightcurve2}
\end{figure}

\begin{figure}
\epsscale{1.0}
\plotone{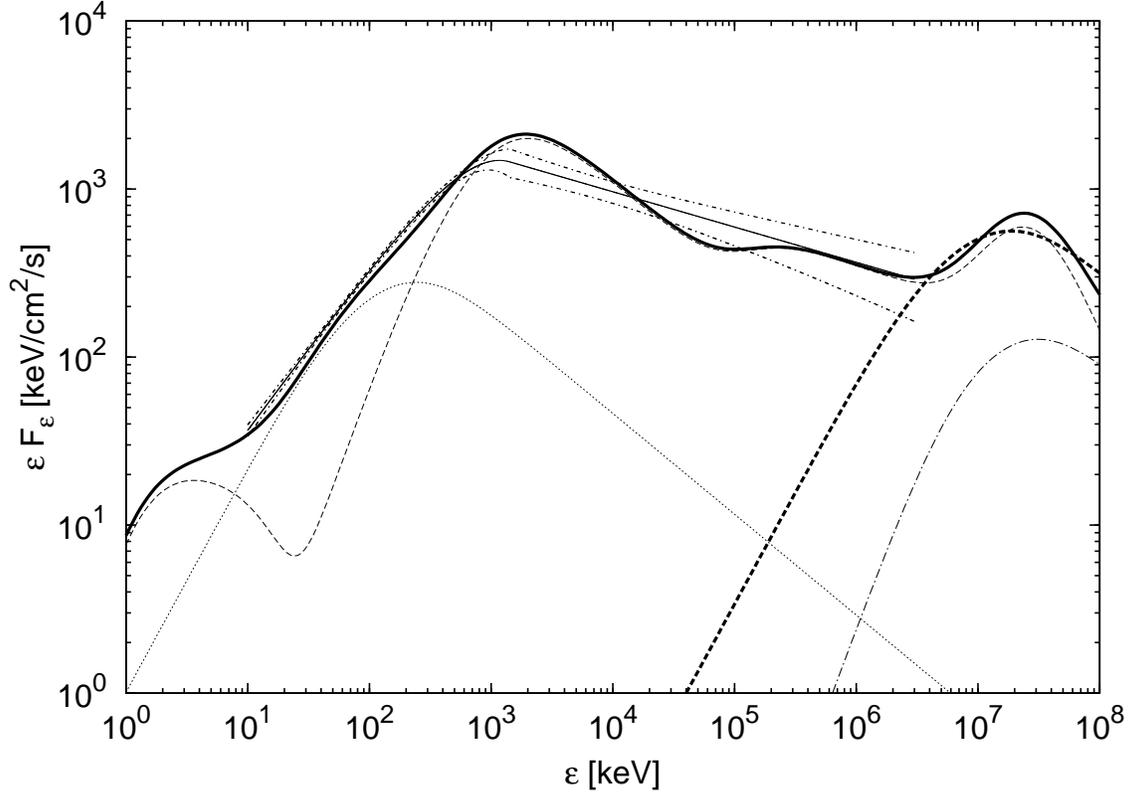}
\caption{
Combined time-averaged emission spectrum of the second and third internal shocks
and the cocoon.
The {\it dashed line} is the same as the thick solid line in 
Figure~\ref{fig:spectrum}, i.e., the cocoon X-rays $+$ the SSC and up-scattered
cocoon (UC) emission of the second internal shock of the jet.
A third less-energetic internal shock occurring at the same observer time as the 
second internal shock but far in front of it produces synchrotron emission 
({\it dotted line}), and this emission is up-scattered into the GeV range by the 
second internal shock shell, arriving at the earth $\simeq 3.2$~s after the
onset of the second pulse ({\it thick dashed line}).
The optical synchrotron emission from the second internal shock, up-scattered 
by the third internal shock electrons, is shown by the {\it dot-dashed line}.
The combined spectrum of the components arriving as the second pulse, i.e.,
the dashed line + the dotted line + the dot-dashed line, is represented by
the {\it thick solid line} and it is consistent with the Band model spectrum 
({\it thin solid line}) with 95\% confidence errors 
({\it dot-short-dashed lines}) (from the LAT/GBM group of {\it Fermi}).
The bump at $\sim 30$~GeV is so dim as not to be detected.
The adopted parameters of the third internal shock are listed in equation 
(\ref{eq:jet_parameter_ad}).
}
\label{fig:spectrumff}
\end{figure}

\clearpage

\appendix

\section{Flux of instantaneous emission from thin shell}
\label{sec:appendix}

We can calculate the observed specific flux from the shell expanding toward us
relativistically by \citep{granot99,woods99}
\begin{equation}
F_\varepsilon (t) = \frac{1+z}{d_L^2} \int d\phi \int d(\cos\theta) \int dr
r^2 \frac{j'_{\varepsilon'}(\mathbf{r},\bar{t})}{\Gamma_j^2(1-\beta_j\cos\theta)^2},
\end{equation}
where $\varepsilon' = (1+z)\varepsilon \Gamma_j(1-\beta_j \cos\theta)$ and 
$t = (1+z)[\bar{t} - (r/c)\cos\theta]$.
For the instantaneous emission from the infinitely thin shell, the comoving 
emissivity can be approximated by \citep[see also][]{ioka01,yamazaki03,dermer04}
\begin{equation}
j'_{\varepsilon'}(\mathbf{r},\bar{t}) \to j'_{\varepsilon'}
\Delta t' \Delta r' \delta(\bar{t} - \bar{t}_i) \delta(r - r_i),
\end{equation}
where $\Delta t' = r_i/c\Gamma_j$ and $\Delta r' = r_i/\Gamma_j$ are
the comoving dynamical timescale and the comoving width of the shell.
The delta functions represent the approximation of the instantaneous emission
at $\bar{t} = \bar{t}_i$ in the infinitely thin shell at $r = r_i$.
The integration can be performed straightforwardly as
\begin{equation}
F_{\varepsilon}(t) = F_{\varepsilon}(t_0) \frac{1}{[1+\Gamma_j^2 \theta^2(t)]^2},
\end{equation}
where 
\begin{equation}
F_{\varepsilon}(t_0) = \frac{1+z}{d_L^2} 8\pi r_i^3 j'_{\varepsilon'}
\end{equation}
\begin{equation}
\varepsilon' = (1+z)\varepsilon \frac{1+\Gamma_j^2 \theta^2(t)}{2\Gamma_j}
\end{equation}
\begin{equation}
\theta(t) = \sqrt{2} \left[1- \frac{c}{r_i} 
\left(\bar{t}_i - \frac{t}{1+z}\right)\right]^{1/2}.
\end{equation}
For the up-scattered cocoon emission, $j'_{\varepsilon'}$ is given by 
equation (\ref{eq:cocoon_emissivity}) and 
this leads to equation (\ref{eq:lightcurve}).

For the 1st-order SSC emission, the comoving emissivity is given by
\begin{equation}
j'_{\varepsilon'} = \tau \frac{\sqrt{3} e^3 B' n'}{4\pi m_e c^2} f(\varepsilon'),
\end{equation}
where
\begin{equation}
f(\varepsilon') = \left\{
\begin{array}{lc}
\left(\frac{{\varepsilon_a^{\rm SC}}'}{{\varepsilon_c^{\rm SC}}'}\right)^{1/3} 
\left(\frac{\varepsilon'}{{\varepsilon_a^{\rm SC}}'}\right)^{1}, 
& {\rm for}~\varepsilon'<{\varepsilon_a^{\rm SC}}', \\
\left(\frac{\varepsilon'}{{\varepsilon_c^{\rm SC}}'}\right)^{1/3},
& {\rm for}~{\varepsilon_a^{\rm SC}}' < \varepsilon' < {\varepsilon_c^{\rm SC}}', \\
\left(\frac{\varepsilon'}{{\varepsilon_c^{\rm SC}}'}\right)^{-1/2},
& {\rm for}~{\varepsilon_c^{\rm SC}}'<\varepsilon'<{\varepsilon_m^{\rm SC}}', \\
\left(\frac{{\varepsilon_m^{\rm SC}}'}{{\varepsilon_c^{\rm SC}}'}\right)^{-1/2} 
\left(\frac{\varepsilon'}{{\varepsilon_m^{\rm SC}}'}\right)^{-p/2},
& {\rm for}~{\varepsilon_m^{\rm SC}}' < \varepsilon',
\end{array}
\right.
\end{equation}
and
\begin{equation}
{\varepsilon_a^{\rm SC}}' = \frac{1+z}{2\Gamma_j}\varepsilon_a^{\rm SC}, ~~
{\varepsilon_c^{\rm SC}}' = \frac{1+z}{2\Gamma_j}\varepsilon_c^{\rm SC}, ~~
{\varepsilon_m^{\rm SC}}' = \frac{1+z}{2\Gamma_j}\varepsilon_m^{\rm SC}.
\end{equation}
This leads to
\begin{equation}
F_{\varepsilon}(t) = \tau F_{\varepsilon_c} 
\frac{f(\nu')}{[1+\Gamma_j^2 \theta^2(t)]^2}.
\end{equation}


\end{document}